\documentclass[a4paper,11pt]{article}
\pdfoutput=1
\usepackage{jheppub}
\usepackage{amssymb,amsmath}
\usepackage{mathtools}
\usepackage{dsfont}
\usepackage{graphicx}
\usepackage{caption}
\usepackage{subcaption}
\usepackage{slashed}
\usepackage{enumerate}
\usepackage{hyperref}
\usepackage{booktabs}

\usepackage{xcolor}
\usepackage{amsmath}

\usepackage{stackengine,scalerel}

\stackMath
\usepackage{float}
\usepackage{placeins}

\newcommand{\nn}{\nonumber}

\title{The Devil Went Down to de Sitter\\[0.3em]
\Large\mdseries Divergent Effective Field Theory Expansions in Cosmology}

\author[a]{Daniele Dorigoni}
\emailAdd{daniele.dorigoni@durham.ac.uk}
\author[b]{Laura Engelbrecht}
\emailAdd{laura.engelbrecht@sns.it}
\author[c]{Nadine Nussbaumer}
\emailAdd{n.a.nussbaumer@rug.nl}
\author[b]{Guilherme L. Pimentel}
\emailAdd{guilherme.leitepimentel@sns.it}

\affiliation[a]{Centre for Particle Theory and Department of Mathematical Sciences, Durham University, Durham DH1 3LE, UK
}
\affiliation[b]{Scuola Normale Superiore and INFN, Piazza dei Cavalieri 7, 56126, Pisa, Italy

}
\affiliation[c]{Van Swinderen Institute for Particle Physics and Gravity,
University of Groningen, Nijenborgh 3, 9747 AG Groningen, The Netherlands}

\abstract{
Effective field theories (EFTs) in de Sitter space, and in accelerating cosmologies more generally, must capture phenomena absent in flat space, most notably long-range correlations generated by spontaneous particle production. When heavy physics is integrated out in such time-dependent backgrounds, these effects leave an imprint on the EFT expansion: even at tree level, this is generically an asymptotic rather than convergent series. This is not a failure of EFT. The large-order behavior of the expansion coefficients encodes physical information inaccessible at any finite order, so cosmological EFTs contain hidden data. Using resurgence analysis, we show how this information can be systematically extracted to recover \textit{exact} and \textit{non-perturbative} predictions from the perturbative expansion. In the example we study, the resurgent completion is naturally organized in terms of the de Sitter $S$-matrix. This leads us to conjecture a prescription for extracting $S$-matrix data from cosmological EFTs in general. The framework extends to any setting where particle production, or a similar non-perturbative phenomenon, renders the EFT expansion asymptotic. 
A longer companion paper will provide a more detailed and pedagogical treatment.}

\begin{document}

\maketitle
\flushbottom


\section{Introduction}
\label{sec:intro}
Effective field theory provides a powerful framework for describing low-energy physics when the details of the underlying high-energy theory are unknown or simply intractable. In flat space, the effects of heavy degrees of freedom are captured by a derivative expansion whose validity is controlled by the ratio of the relevant energy scale to the heavy mass. In cosmological settings, however, time-dependent backgrounds lead to qualitatively different behavior. In particular, as shown in \cite{DuasoPueyo:2025lmq,Li:2026xhw}, the derivative expansion of cosmological correlators generically takes the form of a divergent asymptotic series, even at tree-level. Non-perturbative effects are visible for any kinematics in the physical regime.\footnote{We are always working with comoving, superhorizon momenta, as is standard in inflationary cosmology.} Even near the equilateral limit, where contact self-interactions are expected to dominate the signal, this imprint persists.

Niels Abel famously described divergent series as ``\textit{the Invention of the Devil}''~\cite{Abel:1902}. Nonetheless, they are ubiquitous in physics, where perturbative expansions are
almost always divergent, perhaps most famously exemplified by Dyson's work~\cite{Dyson:1952zz}. By optimally truncating an asymptotic expansion, one can make a deal with the devil, obtaining an approximation whose error is exponentially small in the inverse expansion parameter; see, for example, the review~\cite{Boyd:1999}. But the bargain comes at a price: albeit extremely small, a residual error remains that cannot be further reduced by truncation, suggesting that the EFT expansion cannot exactly capture the full UV completion.

However, \'Ecalle resurgence theory~\cite{Ecalle:1981} provides a
mathematical framework for associating exact expressions to such divergent
series. Crucially, these exact quantities are not determined by perturbation theory
alone: they also contain non-perturbative, exponentially suppressed contributions, which are invisible
at any finite order yet are encoded in the large-order behavior of the perturbative coefficients. In cosmological EFTs, this raises a natural question: can the non-perturbative completion be fully recovered from the information encoded in a divergent EFT expansion? 

In this work, we investigate how the apparent breakdown of the EFT expansion in time-dependent cosmological backgrounds can be systematically understood and treated. We focus on a particular example: the EFT resulting from the exchange of a massive scalar between conformally coupled scalar fields, where the massive scalar has been integrated out. This is a tractable setting in which the EFT expansion and its UV completion can be studied explicitly. Although the UV completion is known in this case, our techniques can also be applied to setups in which the microscopic theory is genuinely unknown. Remarkably, even with incomplete information in the form of a finite number of EFT coefficients, when supplemented by physically motivated constraints, we can reconstruct the non-perturbative effects coming from the UV.  

More specifically, we study the Borel plane associated with the expansion of the correlator at large mass of the exchanged scalar field and fixed kinematics. We use the term ``correlator'' provisionally, since the same perturbative EFT expansion is shared by multiple quantities, and only after identifying the appropriate Borel resummation prescription can we determine which observable is recovered through resurgence. The singularities of the Borel plane encode the presence of non-perturbative effects associated with spontaneous particle production~\cite{Li:2026xhw}. 
A closely related phenomenon occurs in strong-field QED, where the divergence of the derivative expansion of the Euler--Heisenberg effective action is directly tied to non-perturbative Schwinger pair production~\cite{Dunne:1999uy,Florio:2019hzn,Dunne:2022esi}.

Using Borel-Padé approximants, we infer the singularity structure of the Borel transform from a finite set of EFT coefficients. We provide a prescription for selecting a Borel resummation procedure whose analytic structure is compatible with physical requirements and identify the resummed quantity as the `de Sitter $S$-matrix' of \cite{Melville:2023kgd,Melville:2024ove}.  From the $S$-matrix, cosmological observables such as expectation values of field operators and wave function coefficients can be easily extracted by simple integrals in phase space. We expect our prescription for Borel resummation to extend to cosmological correlators in more general theories, providing a route to defining the corresponding $S$-matrix.

Although we use cosmological correlators in the highly symmetric arena of de Sitter space as an example to benchmark the techniques explained in this paper, it is clear that such phenomena are much more general and not tied to specific symmetries of the problem. Rather, they will always be present in scenarios where there is spontaneous particle production from the background geometry~\cite{BirrellDavies:1982}. This includes accelerated cosmologies, gravitational systems with horizons like black holes~\cite{Hawking:1975}, but also thermal field theories, where thermal fluctuations can generate Boltzmann-suppressed correlations at long distances. We conjecture that an analogous resummation prescription exists in all such settings, providing a general framework for recovering non-perturbative effects associated with background-induced particle production or thermal fluctuations.
\vspace{0.2cm}

 While our work was being completed, Y.~Li and Z.-Z.~Xianyu informed us that their work \cite{Li:2026xhw}, exploring similar questions, was being finalized.

\section{Asymptotic EFT Expansions in de Sitter}\label{sec:div}

We consider an EFT expansion of a four-point process involving four conformally coupled\footnote{The mass of a conformally coupled scalar is dimension-dependent. In four spacetime dimensions, $m_\phi^2=2H^2$, while in two spacetime dimensions, $m_\phi^2=0$.} scalar fields in de Sitter spacetime. The expansion is organized in inverse powers of a large parameter $\mu^2$, and takes the form
\begin{equation}\label{eq:FEFT}
   \hat{F}_{\rm EFT}(u,v;\mu)= \overset{\infty}{\underset{n=0}{\sum}}\frac{a_n(u,v)}{\mu^{2n+2}},
\end{equation}
where the coefficient functions $a_n(u,v)$ encode the kinematic dependence at each order in the expansion.\footnote{In the example considered here, $a_n$ depends on two kinematic variables, but in more general settings it may depend on additional variables; see Appendix \ref{app:ds-observables}. } Successive terms capture contributions at progressively higher orders in a derivative expansion.

To illustrate how particle production affects an EFT expansion in de Sitter spacetime and to benchmark our prescription, we study the EFT expansion associated with a tree-level UV completion arising from the $s$-channel exchange of a heavy scalar $\chi$ with mass $m_{\chi}$, coupled to $\phi$ through a cubic interaction of the form $\phi^2 \chi$. We define the dimensionless, large mass parameter $\mu$ to be
\begin{equation} \mu^2\coloneqq\frac{m_{\chi}^2}{H^2}-\frac{d^2}{4}\gg1,
\end{equation}
where $d$ denotes the number of spatial dimensions and $H$ is the Hubble parameter.
The corresponding four-point function $\hat{F}(u,v;\mu)$ depends on the kinematic variables $u$ and $v$, as well as the heavy scalar mass parameter $\mu$. Using bootstrap techniques~\cite{Arkani-Hamed:2018kmz}, the computation of the UV completion can be reduced to solving the differential equation
\begin{equation}\label{diffeq}    \left(\Delta_u+\mu^2\right)\hat{F}(u,v;\mu)=\frac{uv}{u+v},
\end{equation}
where we define\footnote{Notice that the definition of $\Delta_u$ here absorbs the factor of $1/4$ that is present in the mass term of \cite{Arkani-Hamed:2018kmz}.} $\Delta_u\coloneqq u^2(1-u^2)\partial_u^2-2u^3\partial_u+\frac{1}{4}$. This equation is supplemented by physically motivated boundary conditions, which determine the analytic structure of the solution. In particular, the solution in the complex $u$-plane for fixed $v$ in the physical regime $0<v<1$ has the following properties:
\begin{itemize}
    \item $\hat{F}(u,v;\mu)$ is symmetric under the exchange $u \leftrightarrow v$.
    \item $\hat{F}(u,v;\mu)$ is regular in the limit $u\rightarrow 1$, a condition fixed by the choice of the Bunch--Davies vacuum. 
    \item $\hat{F}(u,v;\mu)$ has a branch cut along the real interval $-1\leq u\leq 0$, and is analytic elsewhere in the complex $u$-plane.
\end{itemize} 
The solution to \eqref{diffeq} satisfying the required boundary conditions was derived in \cite{Arkani-Hamed:2018kmz} and is reviewed in Appendix \ref{app:ds-observables}. These properties do not fix the solution uniquely, but rather leave one free parameter, $\beta_0$. Three physical objects commonly studied in this context are the in-in correlator, the wave function coefficient, and the de Sitter $S$-matrix. They all share the same perturbative EFT expansion, satisfy the differential equation \eqref{diffeq}, and exhibit the required analytic structure. Crucially, they correspond to different choices of $\beta_0$, as summarized in Table \ref{tab:objects}. The correlator and wave function coefficient are associated with the process in $D=4$ spacetime dimensions, whereas the de Sitter $S$-matrix considered here corresponds to $D=2$~\cite{Melville:2023kgd,Melville:2024ove}. The $D=4$ de Sitter $S$-matrix can then be related via a simple differential operator in the kinematic variables $u,v$ to its $D=2$ counterpart, see equation \eqref{eq:app-dimension-raising}.
We refer to Appendix \ref{app:ds-observables} for more details.

\begin{table}[!t]
\centering
\renewcommand{\arraystretch}{1.0}
\setlength{\tabcolsep}{6pt}
\begin{tabular}{@{} l @{\hspace{30pt}} c @{}}
    \toprule
   Physical Observable & $\beta_0$ \\
    \midrule
    In-in correlator in $D=4$
        & $\displaystyle\frac{1}{i\sinh(\pi\mu)}$ \\[10pt]
    Wave function coefficient in $D=4$
        & $\displaystyle\frac{R_\mu(s\tau_0)}{i\sinh(\pi\mu)}$ \\[10pt]
    $S$-matrix in $D=2$
        & $0$ \\
    \bottomrule
\end{tabular}\caption{Values of $\beta_0$ corresponding to the three physical observables discussed. The wave function coefficient depends on the late-time regulator $\tau_0$, with $R_\mu$ defined in appendix~\ref{app:ds-observables}.}\label{tab:objects}
\end{table} 

The effective field theory (EFT) expansion arising from the UV completion $\hat{F}(u,v;\mu)$ can be constructed by formally inverting the differential operator $(\Delta_u+\mu^2)$ in equation \eqref{diffeq}, treating $\mu$ as a large parameter, and expanding in inverse powers of  ${\mu^2}$. The resulting EFT then takes the form of \eqref{eq:FEFT} where
the perturbative coefficients are given by
\begin{equation}
a_n(u,v)\coloneqq\left(-\Delta_u\right)^n\left(\frac{uv}{u+v}\right)\,.\label{eq:cuv}
\end{equation}
The resulting EFT describes the four-point process through an infinite tower of higher-derivative contact interactions involving only the conformally coupled scalar.

\begin{figure}[t]
\centering
\includegraphics[width=0.65\textwidth]{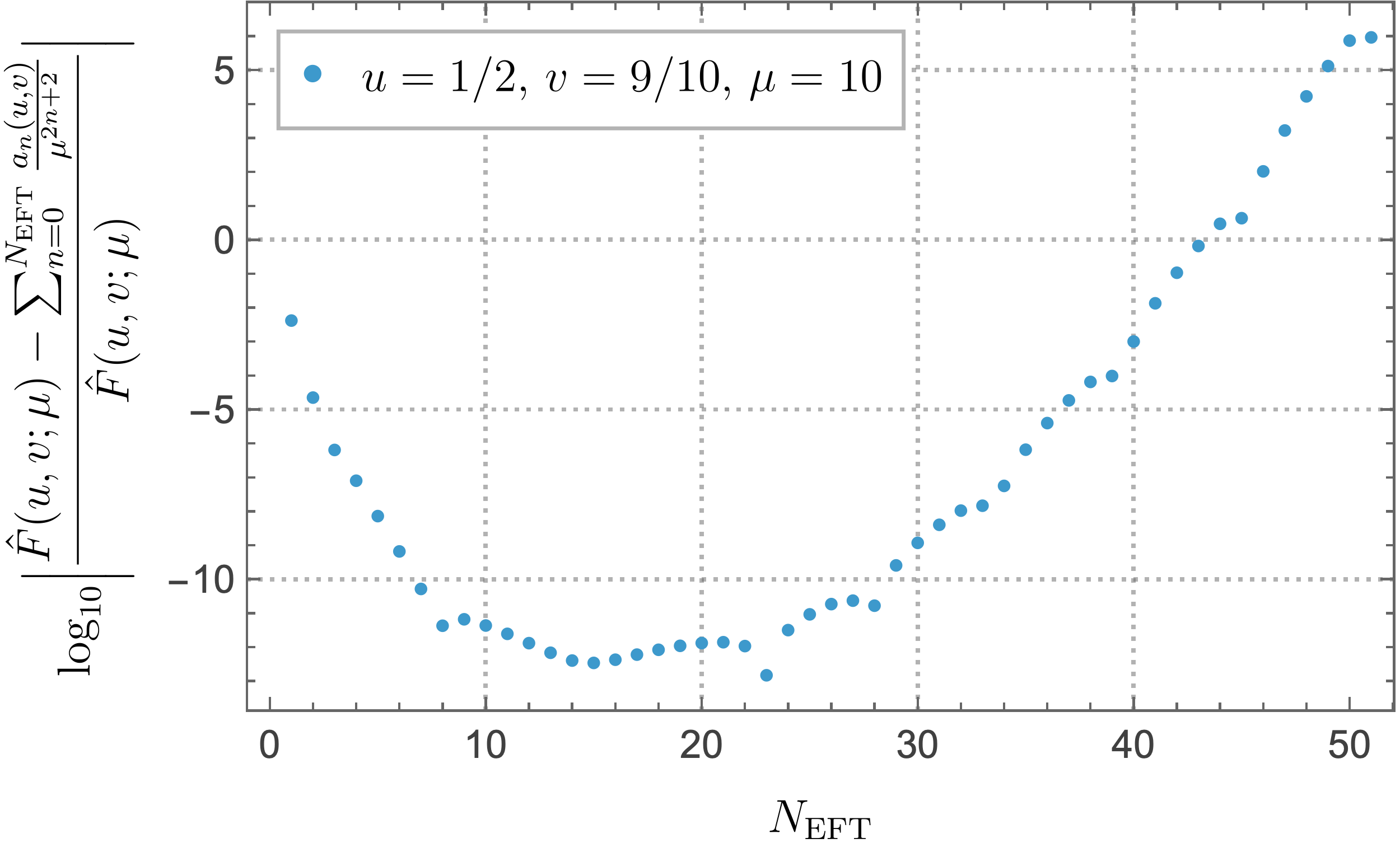}
\caption{Relative error between the dS $S$-matrix and the EFT expansion as a function of the number of perturbative terms $N_{\rm EFT}$ included, at fixed $u=1/2$, $v=9/10$, and $\mu=10$. 
}
\label{fig:EFTPlot}
\end{figure}

The asymptotic nature of the EFT expansion is illustrated in Figure~\ref{fig:EFTPlot}, which shows the relative error between the full de Sitter $S$-matrix and its EFT approximants as a function of the truncation order $N_{{\rm EFT}}$, for fixed kinematic variables and $\mu$.
As successive terms are included, the accuracy initially improves, until the terms reach a minimum size. Beyond this optimal truncation point, the approximation progressively deteriorates as the contributions from higher-order terms increase. This asymptotic behavior persists across all kinematic configurations. Additionally, while the asymptotic behavior is clear for this particular example, any UV completion with a low-energy EFT involving heavy degrees of freedom that have been integrated out will generically lead to divergent EFT expansions at tree-level in time-dependent backgrounds. 

At large-$\mu$, the exact physical quantity $\hat{F}(u,v;\mu)$ is not fully captured by the perturbative EFT expansion. In an attempt to reproduce the full solution $\hat{F}(u,v;\mu)$ solely from the perturbative data~\eqref{eq:FEFT} and the desired analytic structure, we will use resurgence analysis to construct its large-$\mu$ transseries expansion, which will include an infinite tower of non-perturbative effects. 

\section{Non-perturbative Information from the EFT}
\label{sec:NP}

In this section, we analyze the singularity structure of the EFT expansion in the Borel plane and show how this encodes the non-perturbative physics associated with pair production. In particular, we demonstrate that the resummed EFT can reproduce the desired analytic structure in the kinematic variables outlined in Section \ref{sec:div}.

Resurgence analysis\footnote{We refer to \cite{Dorigoni:2014hea,Aniceto:2018bis,Dunne:2025mye} for pragmatic and physically-oriented introductions to resurgence analysis.} determines the possible non-perturbative sectors that can be associated with the asymptotic EFT expansion, but does not, by itself, uniquely determine the physical UV completion. Additional input is required to select the appropriate solution, which in our example is provided by boundary conditions coming from physical requirements. These conditions fix the analytic structure of the physical quantity, and from the perspective of resurgence, corresponds to a particular choice of integration contour in the Borel plane plus a starting point in the kinematic variables.

We investigate how the motion of Borel singularities under analytic continuation in the kinematic variables encodes the analytic structure of the physical quantity. In particular, we show that the required branch cut structure in the physical plane singles out the imaginary axis as a natural contour in the Borel plane. This choice of contour, together with an appropriate starting point in the physical region, provides a concrete prescription for resumming the EFT expansion. We then study the harmonic oscillator limit to determine which physical object is reconstructed by our choice of resummation and use these insights to formulate a  prescription for general cosmological EFTs.

\subsection{Borel Singularity Locations}
\label{sec:Bsing}

Presently we aim not only to solve the specific problem at hand, but also to develop techniques that can be applied more broadly to cosmological EFTs. Hence we primarily focus our attention to presenting numerical methods widely applicable to situations where analytic results may not be feasible. 
A detailed analytical study of the resurgence properties of specific physical observables will be presented in the longer companion paper~\cite{ToAppear}.

It can easily be observed that the perturbative EFT expansion \eqref{eq:FEFT} is an asymptotic series in inverse powers of $\mu^2$ whose perturbative coefficients grow schematically like $a_n(u,v)\sim(2n)!$ . 
This motivates the introduction of the Borel transform
\begin{equation}\label{eq:Borel}
B(u,v;t)\coloneqq\sum_{n=0}^\infty a_n(u,v)\frac{t^{2n}}{(2n)!},
\end{equation}
which has finite radius of convergence around $t= 0$.
It is then possible to resum the formal perturbative expansion \eqref{eq:FEFT} by means of a \textit{directional Laplace transform} of the Borel transform \eqref{eq:Borel}, that is we define
\begin{equation}
\mathcal{S}_\theta[\hat{F}_\textrm{EFT}](u,v;\mu) \coloneqq \frac{1}{\mu}\int_0^{e^{i\theta} \infty} e^{-t\mu} B(u,v;t) \,{\rm d}t\,,\label{eq:Directional}
\end{equation}
where the angle $\theta$ sets the direction of integration in the Borel plane.

The key question is whether there exists a direction of integration $\theta$ (or as we shall see a combination thereof) such that $\hat{F}(u,v;\mu) \stackrel{?}{=}\mathcal{S}_\theta[\hat{F}_\textrm{EFT}](u,v;\mu)$.
If the Borel transform $B(u,v;t)$ has a singular direction, called a \textit{Stokes line}, say for ${\rm arg}\, t = \theta_\star$, then $\mathcal{S}_{\theta_\star\pm\epsilon}[\hat{F}_\textrm{EFT}]$ will produce two different analytic continuations in the variable $\mu$ as $\epsilon\to0^+$. 
Therefore, the singularity structure of $B(u,v;t)$ determines the possible Borel resummations and plays a central role in reconstructing the non-perturbative completion of the EFT.

Studying the singularity structure of the Borel transform \eqref{eq:Borel} is in general difficult, especially when the perturbative coefficients $a_n(u,v)$ are complicated or, as in the present case, depend on external parameters. Such \textit{parametric resurgence} can exhibit a variety of intricate phenomena: singularities in the Borel plane can move, collide, and change their nature as the parameters are varied, which can make them extremely challenging to track analytically, see for example the recent \cite{AnicetoCrew:2026}. Moreover, in an EFT we often have no computationally accessible way of determining $a_n(u,v)$ for all $n\in\mathbb{N}$, and must instead work with a finite number of perturbative coefficients. 

\begin{figure}[t]
  \centering

  \noindent
  \begin{minipage}[t]{\textwidth}
    \begin{subfigure}[t]{0.4\textwidth}
      \centering
      \includegraphics[width=\linewidth]{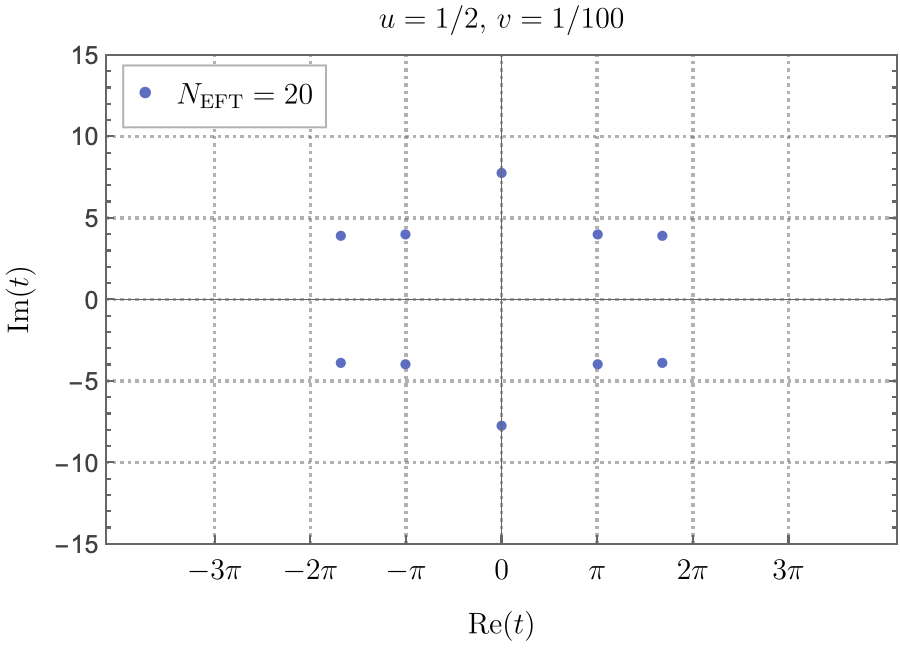}
      \label{fig:im20}
    \end{subfigure}\hspace{14pt}
    \begin{subfigure}[t]{0.4\textwidth}
      \centering
      \includegraphics[width=\linewidth]{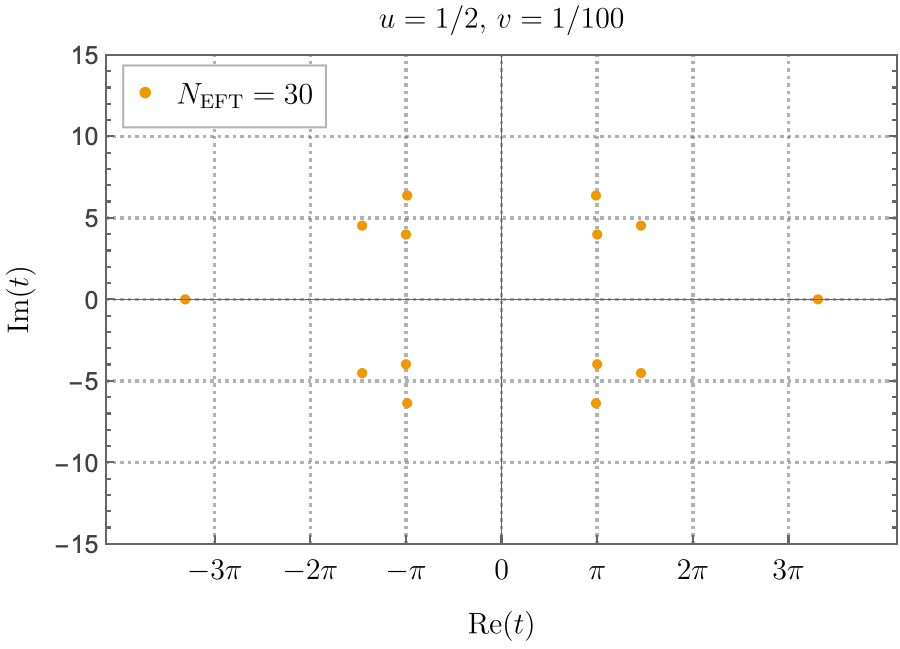}
      \label{fig:im30}
    \end{subfigure}
    \hspace{50pt}
  \end{minipage}


  \noindent
  \begin{minipage}[t]{\textwidth}
    \begin{subfigure}[t]{0.4\textwidth}
      \centering
      \includegraphics[width=\linewidth]{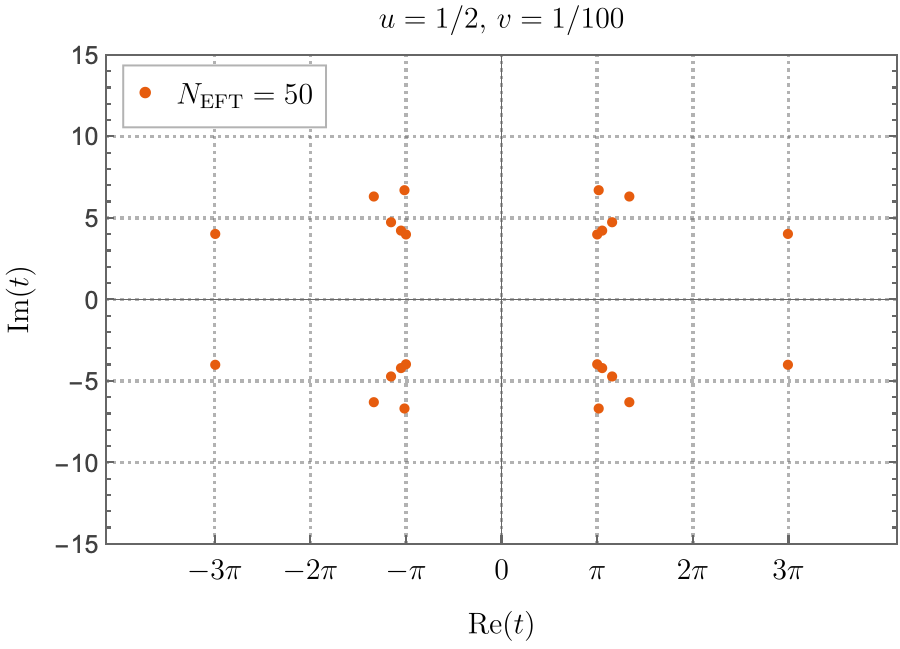}
      \label{fig:im50}
    \end{subfigure}\hfill
    \begin{subfigure}[t]{0.56\textwidth}
      \centering
      \includegraphics[width=\linewidth]{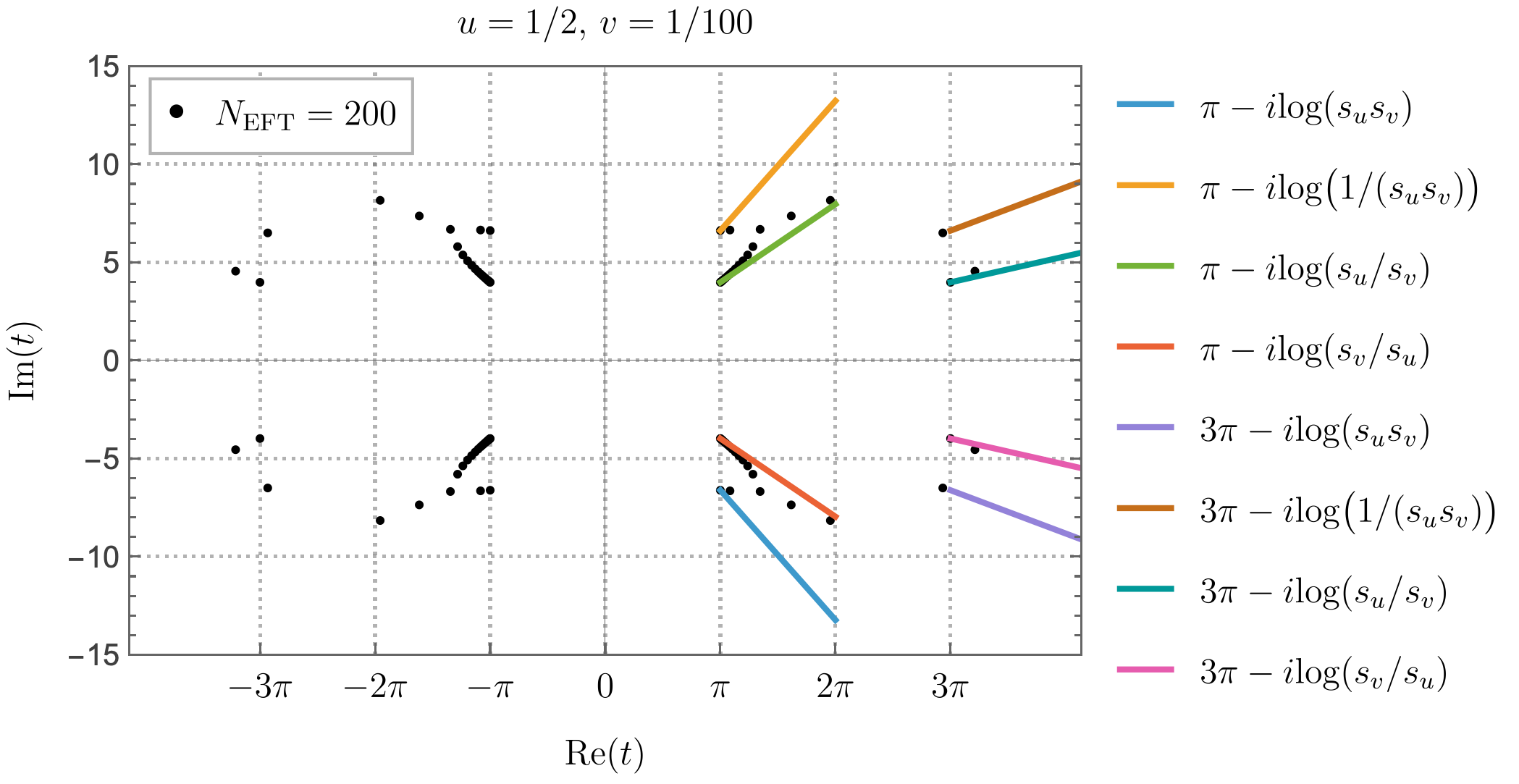}
      \label{fig:im200}
    \end{subfigure}
  \end{minipage}

  \caption{Poles of the diagonal Borel-Pad\'e approximants as we increase the number $N_{\rm EFT}$ of perturbative terms included, starting from $N_{\rm EFT}=20$ up to $N_{\rm EFT}=200$. The branch cuts set by~\eqref{eq:Branches} are marked by the colored lines in the bottom-right plot. }
  \label{fig:borel-plane-panel}
\end{figure}

Importantly, we can identify the singularities of the Borel transform numerically using Padé approximants. Rather than approximating the Borel transform with a finite degree $N_{\rm EFT}$ polynomial, Pad\'e approximants use the same amount of perturbative data to construct rational function approximations for which the degree of the numerator plus the degree of the denominator equals $N_{\rm EFT}$~\cite{BakerGravesMorris:1996}.  The poles of the Padé approximant provide a diagnostic of the analytic structure of the Borel transform. As the order of the approximation is increased, isolated poles that stabilize in position signal genuine pole singularities, while sequences of poles that accumulate along a locus indicate the presence of a branch cut, as neatly depicted in Figure \ref{fig:borel-plane-panel}. This allows us to determine both the location and the type of the leading singularities and to track their motion as the kinematic variables $u$ and $v$ are varied. 

While the poles of Padé approximants provide a useful diagnostic of the singularity structure, the precise arrangement of poles associated with branch cuts requires some care. For diagonal Pad\'e approximants (i.e. rational functions with numerator and denominator of the same degree), Stahl's theorem states
that the poles accumulate on the set of minimal logarithmic capacity connecting
the branch points~\cite{Stahl:1997}. While for a single branch point this is a ray from the branch
point to infinity, for multi-branch cuts, the structure becomes more convoluted.
The approximate cut structure seen in the Borel plane in Figure \ref{fig:borel-plane-panel}  is therefore
one particular choice of branch cuts, not an intrinsic property of the
resurgent function. Any other choice of cuts joining the same branch points defines another
single-valued branch of the same analytic function. Isolated pole-zero pairs that appear and disappear as the
order changes (Froissart doublets) are artifacts of the approximation and should
not be read as singularities. We refer to \cite{BakerGravesMorris:1996} for a thorough discussion of Pad\'e approximants and their properties. 

As a proof of principle, we compute the EFT expansion to sufficiently high order and use Borel-Padé approximants to detect and track the singularity structure in the Borel plane.\footnote{
Note that the numerical accuracy of the approximants can be improved dramatically when additional information is known, such as the precise location and the nature of the singularities of the Borel transform. In particular, conformal maps and uniformizing maps, used together with Pad\'e approximants, provide a systematic way to exploit this information~\cite{Costin:2021bay}.} In practice, we computed $200$ terms of the EFT expansion, constructed the corresponding Borel-Pad\'e approximants, and then studied their pole structure for different kinematic configurations to resolve the leading Borel singularities, as shown in Figure  \ref{fig:borel-plane-panel}. 

For generic $u$ and $v$, we find that the singularity structure is dictated by the quantity
\begin{equation}\label{eq:quant}
s_u \coloneqq \frac{1+\sqrt{1-u^2}}{u}\,,
\end{equation}
and the Borel transform \eqref{eq:Borel} has branch cuts originating at
\begin{equation}\label{eq:Branches}
t_\star = (2n+1)\pi- i \begin{cases} \log(s_u s_v)\,, \\
\log(\frac{1}{s_u s_v})\,,\\
\log(\frac{s_u}{s_v})\,,\\
\log(\frac{s_v}{s_u})\,,
\end{cases}
\end{equation}
with $n\in \mathbb{Z}$.
In the domain $u,v \in(0,1)$, we find that $s_u,s_v>1$, and the singularities of the Borel transform~\eqref{eq:Borel} organize into two families of complex-conjugate branch cuts. The exact analytic structure for the Borel transform \eqref{eq:Borel} can be found in \cite{Li:2026xhw}. 

Several special points in kinematic space are worth noting. At $u=v=\pm1$, the logarithmic contributions in \eqref{eq:Branches} vanish, the four branch cuts associated with each odd multiple of $\pi$ coincide, and the corresponding branch cuts collapse into simple poles on the real Borel axis located at
\begin{equation}
t_\star=(2n+1)\pi.
\end{equation} 
When $u=v$, the equality $s_u=s_v$ implies that two of the branch points coincide. In this case, a series of corresponding cuts lie along the real axis, beginning at each odd multiple of $\pi$. Finally, for $u=-v$, corresponding to the total energy singularity~\cite{Arkani-Hamed:2018bjr}, the EFT coefficients themselves become singular and the expansion considered here is not well-defined.

\subsection{Borel Singularity Trajectories and Analytic Structure of the Physical Plane}

In addition to selecting a particular integration contour, we must also specify a starting point $(u_0,v_0)$ in kinematic space and from there track how the Borel singularities move as the kinematic variables are varied. This is essential for determining the analytic structure of the resummed function in kinematic space. As the kinematic variables are analytically continued, Borel singularities can approach or cross the prescribed integration contour, requiring the contour to be deformed in order to define the resummation continuously. The resummed function itself can become multi-valued as a function of the kinematic variables if approaching a given point from different directions leads to different values.

For example, let us fix $v$ and choose  ${\rm arg}\,t=\theta$ as the contour of integration in the Borel plane. Suppose there exists a value $u_\star$, such that as $u$ approaches $u_\star$, the branch cuts in the Borel plane approach the contour of integration differently depending on whether ${\rm arg}(u/u_\star)\to0^\pm$. This will result in non-trivial branch structures for the directional Borel resummation as function of the kinematic variable $u$, characterized by 
\begin{equation}
\lim_{\epsilon\to0^+}\mathcal{S}_\theta[\hat{F}_\textrm{EFT}](u_\star e^{i\epsilon} ,v;\mu) \neq\lim_{\epsilon\to0^+}\mathcal{S}_\theta[\hat{F}_\textrm{EFT}](u_\star e^{-i\epsilon} ,v;\mu)\,,
\end{equation}
thereby indicating the presence of a singularity in $u$ along the direction ${\rm arg} \, u_\star$. 

 This is the mechanism by which the motion of Borel singularities can generate branch cut structure in the physical variables. Our goal is therefore to determine how the Borel-plane singularities and their motion encode the analytic structure of the resummed function in kinematic space.

We focus on the $u$-dependence, keeping $v$ fixed in the physical regime, $0<v<1$. The physical quantity we wish to reconstruct is analytic in the complex $u$-plane except for a branch cut along the negative real axis, $-1\le u \le 0$, as discussed in Section \ref{sec:div}. For fixed $v$, the dependence of Borel singularities on $u$, therefore, enters through the logarithm of $s_u$ as can be seen from \eqref{eq:Branches}. As $u$ is analytically continued, the real and imaginary parts of these logarithms change, causing the Borel singularities to move in the complex $t$-plane. We analyze these trajectories and determine the regions of the $u$-plane in which the singularities exhibit qualitatively distinct behavior.

\begin{figure}[t]
  \centering
  \begin{subfigure}[t]{0.4\textwidth}
    \centering
    \includegraphics[width=\textwidth]{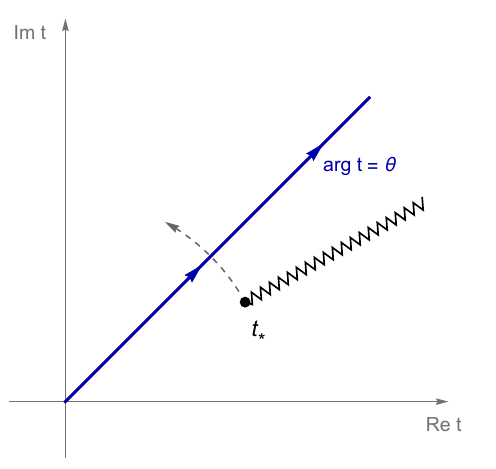}
    \caption{Before crossing.}
    \label{fig:contour-before}
  \end{subfigure}\hspace{0.07\textwidth}
  \begin{subfigure}[t]{0.4\textwidth}
    \centering
    \includegraphics[width=\textwidth]{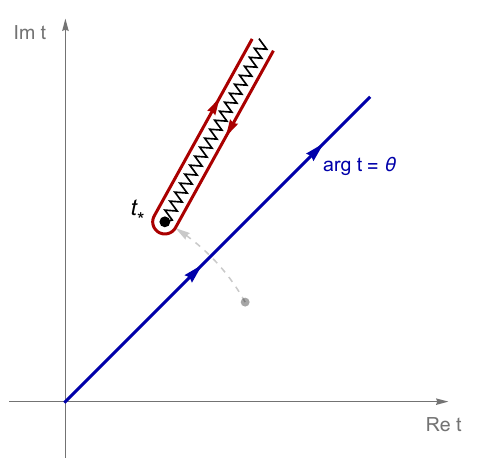}
    \caption{After crossing:  
    Ray $+$ Hankel contour.}
    \label{fig:contour-after}
  \end{subfigure}
  \caption{As we change the kinematic variable $u$, a singular direction can cross the contour of integration in the $t$-Borel plane. The resulting Hankel contour, given in red in (b), produces a purely non-perturbative contribution.}
  \label{fig:contour-crossing}
\end{figure}

What matters is not simply where the singularities are located at a fixed point in kinematic space, but whether their trajectories can reach a candidate integration contour given a specified starting point. As schematically depicted in Figure \ref{fig:contour-crossing}, when a singularity passes through an integration contour, our resummed expression will acquire an additional contribution.
The novel term associated with the Hankel contour circling the branch cut, given by the red contour in Figure \ref{fig:contour-after}, generates non-perturbative, exponential terms of the form
\begin{align}\label{eq:expterm}
\lim_{\epsilon\to0^+}\left[\mathcal{S}_{\theta_\star +\epsilon}-\mathcal{S}_{\theta_\star-\epsilon}\right][\hat{F}_\textrm{EFT}](u,v;\mu) &\notag = \frac{1}{\mu}\int_\gamma e^{-t\mu} B(u,v;t) \,{\rm d}t\\
& = \frac{1}{\mu} e^{-t_\star \mu}\int_0^{e^{ i \theta_\star}\infty } e^{ -t \mu} {\rm disc}_{\theta_\star} B(u,v;t+t_\star) {\rm d}t\,,
\end{align}
where $\theta_\star = {\rm arg}\, t_\star$ and ${\rm disc}_{\theta_\star} B(u,v;t)$ denotes the discontinuity of the Borel transform along the singular direction of the branch cut.

\begin{figure}[t]
  \centering
  \begin{subfigure}[t]{0.49\textwidth}
    \centering
    \includegraphics[width=\textwidth]{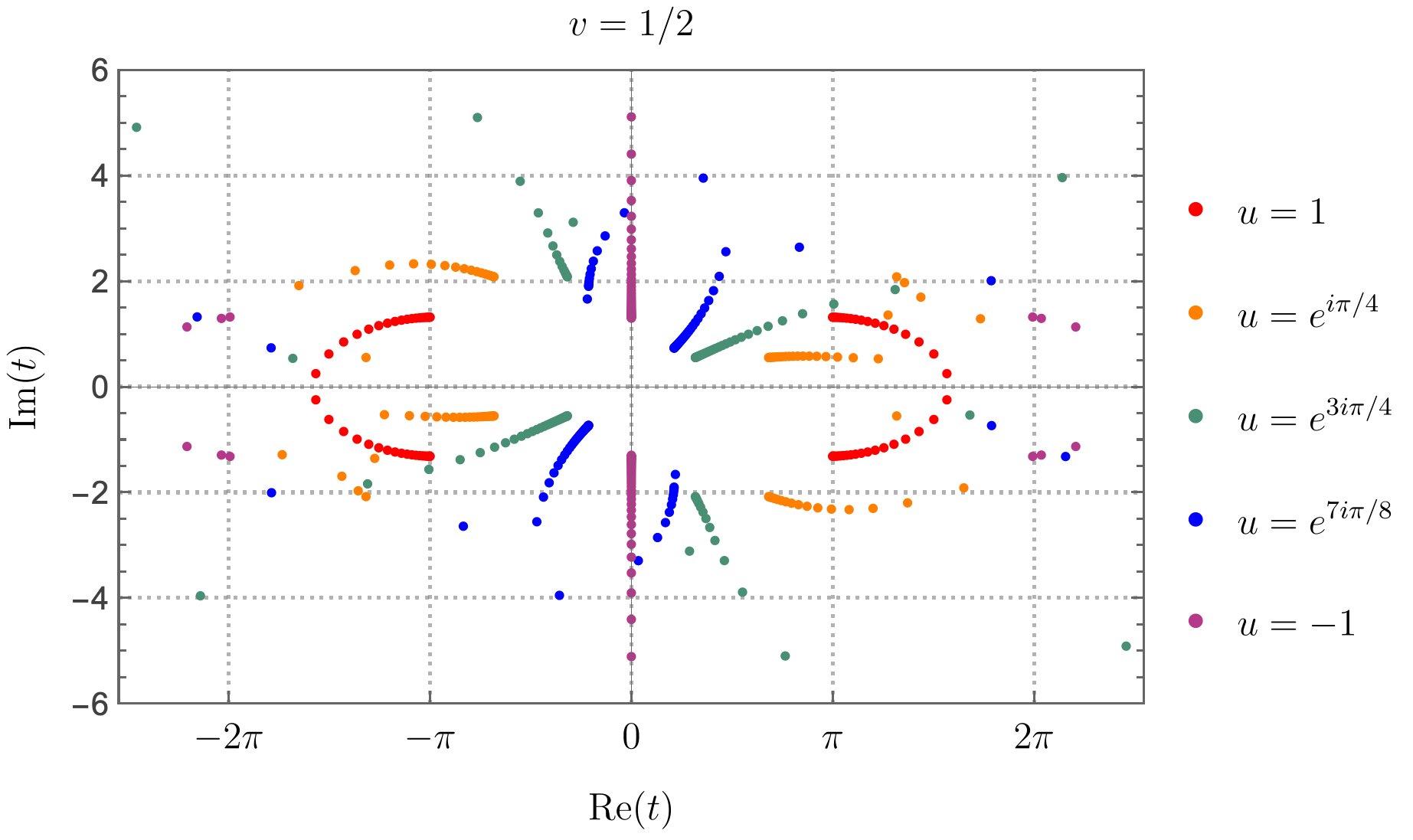}
    \caption{Singularity structure in the Borel $t$-plane as we rotate $u$ along a circle of radius $|u|=1$.}
    \label{fig:fromabove}
  \end{subfigure}\hfill
  \begin{subfigure}[t]{0.49\textwidth}
    \centering
    \includegraphics[width=\textwidth]{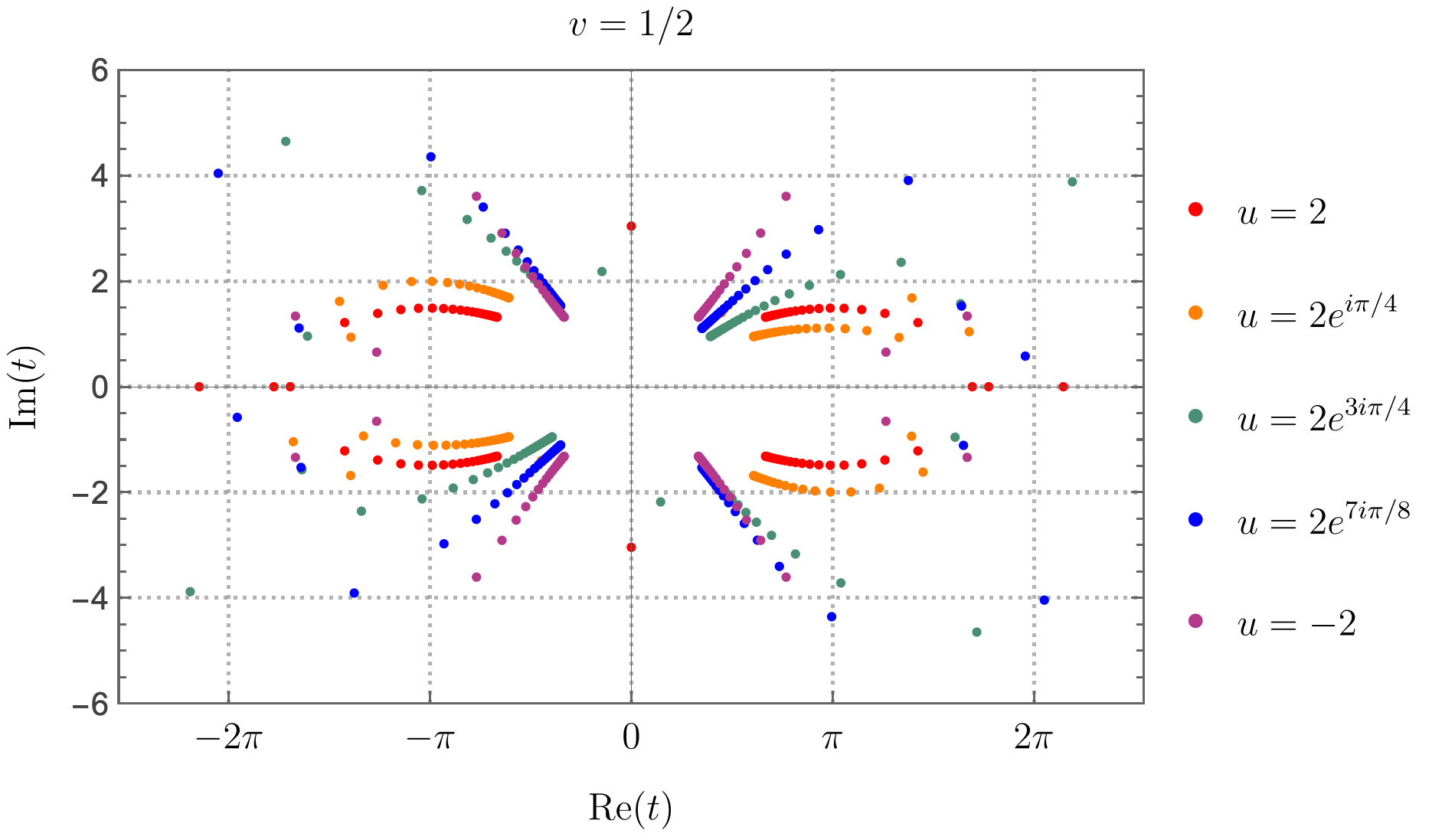}
    \caption{Singularity structure in the Borel $t$-plane as we rotate $u$ along a circle of radius $|u|=2$.}
    \label{fig:fromabover2}
  \end{subfigure}
  \caption{Motion of the Borel singularities as we vary the kinematic variable $u=|u|e^{i\theta}$ at fixed $v=1/2$. For $|u|=1$, a branch cut approaches the imaginary $t$-axis for ${\rm arg}\, u \to\pi$, as shown in (a). For $|u|=2$, no singularity reaches the imaginary $t$-axis as shown in (b).}
  \label{fig:urotation}
\end{figure}

More importantly, the way in which the contour encounters singularities can induce non-trivial branch structure in the kinematic variables.  Since the physical solution is required to develop a branch cut along the interval $-1\leq u\leq0$, we study the behavior of the singularities as $u$ approaches this interval. Tracking the real part of $t_\star$ reveals a striking correspondence: near this interval in the $u$-plane, certain Borel singularities approach and reach the imaginary $t$-axis,
\begin{equation}
\operatorname{Re}t_\star=0.
\end{equation}
Thus, the imaginary axis emerges as a distinguished locus in the Borel plane with singularities reaching it precisely where the physical solution requires non-trivial behavior. In particular, some of the branches closest to the origin, corresponding to $n=-1$ and $n=0$ in \eqref{eq:Branches}, reach the imaginary axis for $u\in [-1,0]$,  while all other branches remain displaced from it, as shown in Figure \ref{fig:urotation}.

To see how the Borel singularities behave as $u$ approaches the branch cut, consider first approaching the negative real axis from above with
\begin{equation}
u=|u|e^{i\theta},\qquad
0<|u|< 1,\qquad
\theta=\pi-\epsilon,\qquad
\epsilon\rightarrow0^+.
\end{equation}
In this limit,
\begin{equation}
\log s_u
=
\log s_{|u|}
-i\pi+i\epsilon f_u,
\end{equation}
where we define
\begin{equation}\label{eq:fu}
f_u\coloneqq\frac{1}{\sqrt{1-|u|^2}},
\end{equation}
and note that $f_u>0$.
The four singularities whose real parts approach zero are then
\begin{equation}\label{eq:revanishabove}
t_\star = \begin{cases} \phantom{-}\epsilon f_u-i\log (s_{|u|}s_v)\,, \\
\phantom{-}\epsilon f_u-i\log(\frac{s_{|u|}}{ s_v})\,,\\
-\epsilon f_u-i\log (\frac{1}{s_{|u|}s_v})\,,\\
-\epsilon f_u-i\log(\frac{s_v}{s_{|u|}})\,.
\end{cases}
\end{equation}

We note that \eqref{eq:fu} is divergent for $|u|=1$. A similar analysis for $u = e^{i(\pi-\epsilon)}$ yields instead
\begin{equation}\label{eq:revanishabove1}
t_\star = \begin{cases} \phantom{-}(1-i) \sqrt{\epsilon }-i\log (s_v)\,, \\
\phantom{-}(1-i) \sqrt{\epsilon }-i\log(\frac{1}{ s_v})\,,\\
-(1-i) \sqrt{\epsilon }-i\log (\frac{1}{s_v})\,,\\
-(1-i) \sqrt{\epsilon }-i\log(s_v)\,.
\end{cases}
\end{equation}

Conversely, approaching the negative real axis from below with
\begin{equation}
u=|u|e^{i\theta},\qquad
0<|u|< 1,\qquad
\theta=-\pi+\epsilon,\qquad
\epsilon\rightarrow0^+,
\end{equation}gives
\begin{equation}
\log s_u
=
\log s_{|u|}
+i\pi-i\epsilon f_u,
\end{equation}and the corresponding singularities are
\begin{equation}\label{eq:revanishbelow}
t_\star = \begin{cases} \phantom{-}\epsilon f_u-i\log (\frac{1}{s_{|u|}s_v})\,, \\
\phantom{-}\epsilon f_u-i\log(\frac{s_{v}}{ s_{|u|}})\,,\\
-\epsilon f_u-i\log (s_{|u|}s_v)\,,\\
-\epsilon f_u-i\log(\frac{s_{|u|}}{ s_v})\,.
\end{cases}
\end{equation}
As before, the above expression is singular for $|u|=1$, thus
repeating the analysis for $u = e^{i (-\pi +\epsilon)}$ yields
\begin{equation}\label{eq:revanishbelow1}
t_\star = \begin{cases} \phantom{-} (1+i) \sqrt{\epsilon }-i\log (\frac{1}{s_v})\,, \\
\phantom{-}(1+i) \sqrt{\epsilon }-i\log(s_{v})\,,\\
-(1+i) \sqrt{\epsilon }-i\log (s_v)\,,\\
-(1+i) \sqrt{\epsilon }-i\log(\frac{1}{ s_v})\,.
\end{cases}
\end{equation}

\begin{figure}[t]
  \centering
  \begin{subfigure}[t]{0.49\textwidth}
    \centering
    \includegraphics[width=\textwidth]{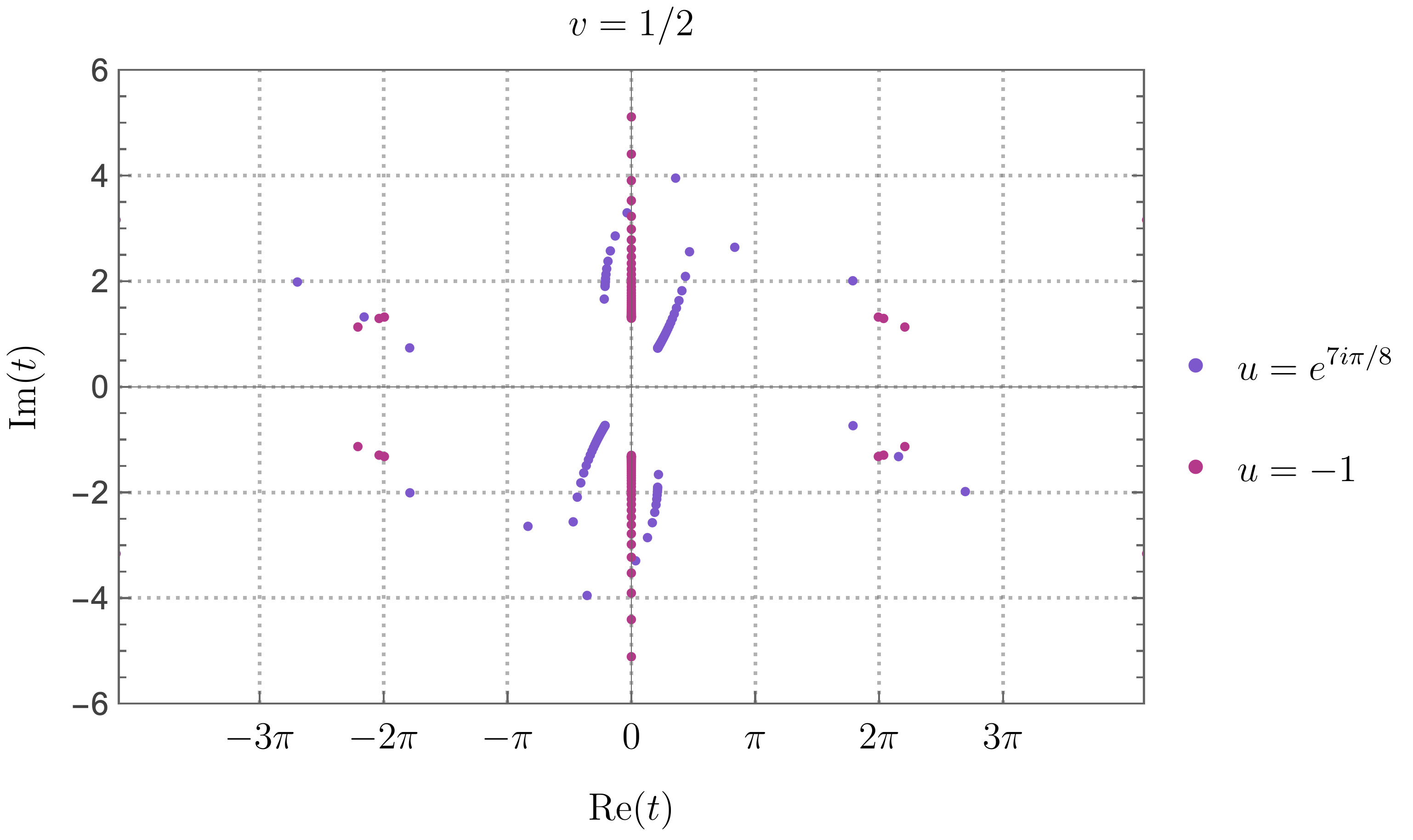}
    \caption{Singularity structure  for $|u|=1$. In magenta ${\rm arg}\, u =\pi$, and in purple ${\rm arg} \,u =\pi(1-\frac{1}{8})$. }
    \label{fig:rotationabove}
  \end{subfigure}\hfill
  \begin{subfigure}[t]{0.49\textwidth}
    \centering
    \includegraphics[width=\textwidth]{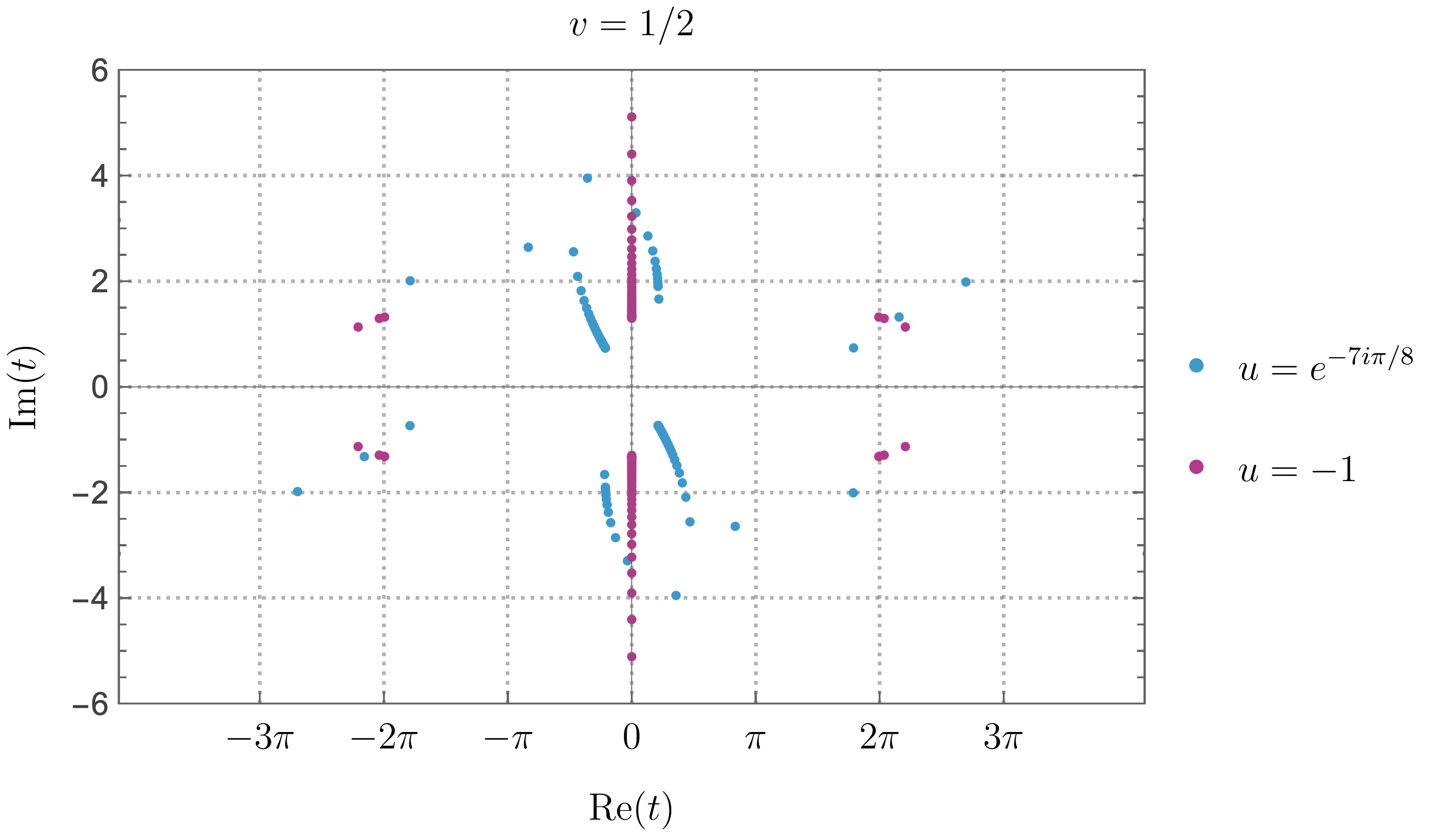}
    \caption{Singularity structure  for $|u|=1$. In magenta ${\rm arg}\, u =-\pi$, and in blue ${\rm arg} \,u =\pi(-1+\frac{1}{8})$.}
    \label{fig:rotationbelow}
  \end{subfigure}
  \caption{Singularity structure of the Borel transform for $u=e^{ i\theta}$ along the unit circle, with $v=\frac{1}{2}$ held fixed. Branch cuts approach the imaginary axis from different directions depending on the limiting value of ${\rm arg}\,u$: ${\rm arg}\,u\to \pi^-$ in (a) or ${\rm arg}\,u\to (-\pi)^+$ in (b).}
  \label{fig:uclose}
\end{figure}

Comparing~\eqref{eq:revanishabove}-\eqref{eq:revanishabove1} and~\eqref{eq:revanishbelow}-\eqref{eq:revanishbelow1}, we see that as ${\rm arg}\,u\to \pi$ with $|u|\leq 1$ we have \textit{different} Borel singularities approaching the imaginary axis: singularities with the same imaginary part approach the axis ${\rm Re} \,t=0$ from opposite sides depending on whether ${\rm arg}\,u=\pi$ is approached from above or below, as illustrated in Figure~\ref{fig:uclose}.

This result is significant because the interval $u\in[-1,0]$ is precisely the region along which the physical solution is required to develop a non-trivial branch structure. The required branch structure in the kinematic variables singles out the imaginary axis as the natural contour. We therefore prescribe the contour of our directional Laplace transform to run along the imaginary $t$-axis, with $u$ and $v$ in the physical regime which we shall denote by $0< u_0,v_0 \le 1$ for its starting point, that is we define
\begin{align}
\mathcal{S}_{\rm avg}[\hat{F}_{\textrm{EFT}}](u_0,v_0;\mu) &\notag\coloneqq \frac{1}{2}\left(\mathcal{S}_{\frac{\pi}{2}} +\mathcal{S}_{-\frac{\pi}{2}}\right) [\hat{F}_{\textrm{EFT}}](u_0,v_0;\mu)\\
&=\frac{1}{2\mu}\left(\int_0^{i \infty} e^{-t\mu} B(u_0,v_0;t) \,{\rm d}t+\int_0^{-i \infty} e^{-t\mu} B(u_0,v_0;t) \,{\rm d}t\right)\,.\label{eq:contour}
\end{align}
 The resummed solution at other values of $(u,v)$ is then obtained by analytically continuing from this starting point while continuously deforming the contour to capture moving Borel singularities it encounters.

The detailed analysis of the trajectories of the Borel singularities as functions of $u$ and $v$ is given in Appendix \ref{app:trajectories}. Here we emphasize the resulting picture: the analytic structure in the kinematic variables is reflected in the motion of singularities in the Borel plane. The singularity trajectories determine where non-trivial structure can arise, while the choice of contour determines how that structure is realized in the resummed solution. 

We have reproduced the necessary analytic structure of the physical quantity, which constrains the resummed EFT to lie within the one-parameter family of solutions given in equation \eqref{fullsol}. However, once the contour and its starting point are specified, there is no remaining freedom in the physical quantity reconstructed by the resummation. Hence the parameter $\beta_0$ has been fixed indirectly.\footnote{{As we will argue for analytically in the longer companion paper \cite{ToAppear}, amongst the values of $\beta_0$ given in Table \ref{tab:objects} the only one compatible with the singularity structure of Section \ref{sec:Bsing} is that of the dS $S$-matrix.}}

Determining which member of this family is selected can be challenging numerically. 
We therefore turn to the harmonic oscillator limit, where the equations simplify and the exact solution can easily be obtained analytically. This provides a controlled setting in which we can determine which physical quantity is reconstructed by our resummation prescription.

\subsection{Harmonic Oscillator Limit} 

As shown in \cite{Arkani-Hamed:2018kmz}, we can obtain a simple harmonic oscillator (SHO) limit by setting $u=\xi v$,  $\hat{F}=v F$, and then taking the limit $v\rightarrow 0$, thus reducing the differential equation \eqref{diffeq} to that of a forced harmonic oscillator
\begin{equation}  
\left[\xi^2\partial_\xi^2+\left(\mu^2+\frac{1}{4}\right)\right]F(\xi;\mu)=\frac{\xi}{\xi+1}.\label{eq:SHOdiffeq}
\end{equation}
The above differential equation admits the general solution
\begin{equation}\label{eq:F}
F(\xi;\mu)  = F_P(\xi;\mu) + C_1(\mu) \xi^{\frac{1}{2} + i\mu }+C_2(\mu)\,  \xi^{\frac{1}{2}-i\mu }\,,
\end{equation}
where $C_1,C_2$ are functions of $\mu$ only and parametrize the two independent solutions to the homogeneous problem, while $F_P$ denotes the particular solution
\begin{equation}
F_P(\xi;\mu)  \coloneqq \frac{i\xi}{\mu(1+2i\mu)} \,_2F_1(\tfrac{1}{2}+i\mu, 1 ; \tfrac{3}{2} +i \mu \vert - \xi)- \frac{i\xi}{\mu(1-2i\mu)} \,_2F_1(\tfrac{1}{2}-i\mu, 1 ; \tfrac{3}{2} -i \mu \vert - \xi)\,.\label{eq:FP}
\end{equation}

As in the full problem, the differential equation \eqref{eq:SHOdiffeq}   immediately gives rise to an asymptotic, perturbative EFT expansion at large $\mu$, given by
\begin{equation}
F_{\textrm{EFT}}(\xi;\mu)=\frac{1}{\Delta_\xi+\mu^2}\left(\frac{\xi}{\xi+1}\right)
=\sum_{n=0}^\infty \frac{b_n(\xi)}{\mu^{2n+2}}\,\,,\qquad b_n(\xi)\coloneqq (-1)^n\Delta_\xi^n\!\left(\frac{\xi}{\xi+1}\right)\,,
\label{eq:FPert}
\end{equation}
with $\Delta_\xi \coloneqq \xi^2 \partial_\xi^2 +1/4$.
The perturbative coefficients $b_n(\xi)$ are of course directly related to the more general $a_n(u,v)$ defined in \eqref{eq:cuv} via $b_n(\xi) = \lim_{v\to0} a_n(\xi v,v)/v$.

Once more, the perturbative EFT coefficients  grow factorially fast $b_n(\xi)\sim (2n)!$. This motivates us to consider a Borel transform analogous to that given in equation \eqref{eq:Borel},
\begin{equation}
B(\xi;t)\coloneqq\sum_{n=0}^\infty b_n(\xi)\,\frac{t^{2n}}{(2n)!}\,.
\label{eq:Bdef}
\end{equation}
In the SHO limit, the Borel transform can be  expressed in closed form as
\begin{align}
B(\xi;t)
=\frac12\left[\frac{\xi e^{\frac{it}{2}}}{1+\xi e^{it}}+\frac{\xi e^{-\frac{it}{2}}}{1+\xi e^{-it}}\right]\,,
\label{eq:Btwoterm}
\end{align}
from which it is straightforward to derive its singularity structure consisting of simple poles located at
\begin{equation}
t_\star^\pm(\xi) \coloneqq (2n+1)\pi \pm i \log(\xi)\,.\label{eq:poles}
\end{equation}
Note that these poles correspond precisely to the two general singularity families from \eqref{eq:Branches} which remain finite in the limit $v\to 0$, i.e.
\begin{equation}
\label{eq:Branches1}
t_\star = (2n+1)\pi- i \begin{cases} \log(\frac{s_{\xi v}}{s_{v}})\,, \\
\log(\frac{s_{v}}{s_{\xi v} })\,,
\end{cases}
\stackrel{v\to 0}{\xrightarrow{\hspace{1cm}}}\begin{cases} t^+_\star(\xi)\,, \\
t^-_\star(\xi)\,.
\end{cases}
\end{equation}

 The physical regime for $\xi$,  which we shall denote by $\xi_0$, is given by $\xi_0\in(0,\infty)$. For $0<\xi_0<1$, the quantity $\log\xi_0<0$, so the $t^+_\star(\xi_0)$ poles lie in the lower half-plane, while the $t^-_\star(\xi_0)$ poles lie in the upper half-plane. As $\xi_0$ approaches $1$ from below, the poles approach the real line and collide at odd multiples of $\pi$. As $\xi_0$ increases beyond $1$, the poles exchange sides of the real axis: the $t^+_\star(\xi_0)$ poles move into the upper half-plane, while the $t^-_\star(\xi_0)$ poles move into the lower half-plane. Thus, 
\begin{align}
\operatorname{Im}t_\star^+(\xi_0)  &\label{eq:sing1}> 0 \,,\qquad \operatorname{Im}t_\star^-(\xi_0) <0\,, \qquad \xi_0>1\,,\\\operatorname{Im}t_\star^+(\xi_0)  &\label{eq:sing}= 0 \,,\qquad \operatorname{Im}t_\star^-(\xi_0) =0\,, \qquad \xi_0=1\,,\\
\operatorname{Im}t_\star^+(\xi_0) \label{eq:sing2}&< 0 \,,\qquad \operatorname{Im}t_\star^-(\xi_0)  >0\,, \qquad 0<\xi_0<1\,.
\end{align}

We now apply the contour prescription given in equation \eqref{eq:contour}. Although the Borel singularities exchange sides of the real axis as $\xi_0$ passes through $1$, the prescription allows us to choose any starting point $\xi_0$ in the physical regime $\xi_0>0$. The resulting contour prescription is illustrated in Figure~\ref{fig:SHOcontours}. A detailed derivation of the resummation is given in Appendix \ref{app:HOresummation}. Performing the resummation, we obtain
\begin{equation}\label{eq:UV-HO}
    \mathcal{S}_{\rm avg}[F_{\textrm{EFT}}](\xi_0;\mu)=F_P(\xi_0;\mu)+\frac{i\pi}{4\mu\cosh\pi\mu}\left(\xi_0^{\frac{1}{2}+i\mu}-\xi_0^{\frac{1}{2}-i\mu}\right).
\end{equation}

We now compare \eqref{eq:UV-HO} with the harmonic oscillator limit of the full solution given in \eqref{fullsol}.  Taking the $u_0<v_0$ case of \eqref{fullsol}, computing the harmonic oscillator limit, and evaluating it at the physical starting point $(\xi_0v_0,v_0)$, the expression reduces to
\begin{align}\label{eq:fullsolho}
   \lim_{v_0\rightarrow0}\, \frac{1}{v_0}\hat{F}(\xi_0 v_0,v_0;\mu)=F_P(\xi_0;\mu)+\frac{i\pi}{4\mu\cosh \pi\mu}\bigg(&\xi_0^{\frac{1}{2}+i\mu}-\xi_0^{\frac{1}{2}-i\mu}+\beta_0\left(\xi_0^{\frac{1}{2}+i\mu}+\xi_0^{\frac{1}{2}-i\mu}\right)\nn\\
   &+h(\xi_0,v_0;\mu)\bigg)\,,
\end{align}
where the function $h(\xi,v;\mu)$ contains dependence on $v$ and is given by
\begin{equation}
   h(\xi_0,v_0;\mu)\coloneqq-\frac{\alpha_-}{\alpha_+}(\beta_0+1)\left(\frac{iv_0
   }{2\mu}\right)^{2i\mu}\xi_0^{\frac{1}{2}+i\mu}-\frac{\alpha_+}{\alpha_-}(\beta_0-1)\left(\frac{iv_0
   }{2\mu}\right)^{-2i\mu}\xi_0^{\frac{1}{2}-i\mu}.
\end{equation}
The quantities $\alpha_\pm$ are functions only of $\mu$ given explicitly in \eqref{eq:alphapm}.
The terms contained in $h$ arise from two of the Borel singularities \eqref{eq:Branches} located at
\begin{equation}
\label{eq:Branches2}
t_\star = (2n+1)\pi- i \begin{cases} \log(s_{\xi_0 v_0} s_{v_0})\,, \\
\log(\frac{1}{s_{\xi_0 v_0} s_{v_0}})\,,
\end{cases}
\end{equation} 
which are driven to infinity in the harmonic oscillator limit $v_0\to 0 $. These contributions therefore decouple from the finite Borel plane and can be neglected in this limit. Comparing the remaining terms in \eqref{eq:fullsolho}
with the resummed result \eqref{eq:UV-HO}, we find that the resummation prescription fixes
\begin{equation}
    \beta_0=0.
\end{equation} 
This identifies the resummed solution \eqref{eq:contour} with the de Sitter $S$-matrix. Thus, in the harmonic oscillator limit, the contour and starting point prescription uniquely selects the de Sitter $S$-matrix from the family of solutions with the required analytic structure.

\subsection{Contour Prescription Conjecture}

The SHO limit allows us to determine which physical quantity is constructed by the resummation procedure in \eqref{eq:contour}. Since $\beta_0$ parametrizes the homogeneous solution of the full differential equation, the same value must also hold when taking the harmonic oscillator limit. The value $\beta_0=0$ obtained in this limit therefore fixes the corresponding parameter in the full theory. We can thus identify the resummation prescription for the cosmological EFT 
\begin{equation}
\mathcal{S}_{\rm avg}[\hat{F}_{\textrm{EFT}}](u_0,v_0;\mu) = \frac{1}{2\mu}\left(\int_0^{i \infty} e^{-t\mu} B(u_0,v_0;t) \,{\rm d}t+\int_0^{-i \infty} e^{-t\mu} B(u_0,v_0;t) \,{\rm d}t\right)\,,
\end{equation}
for $0<u_0,v_0\leq 1$, with the de Sitter $S$-matrix.
The resummed solution at generic $(u,v)$ is then obtained by analytically continuing from $(u_0,v_0)$ while continuously deforming the contour as the Borel singularities move. For the example studied here, this prescription reconstructs the de Sitter $S$-matrix at generic kinematics $(u,v)$. 

As previously remarked, although the above resummation is defined for the physical range of kinematics $0<u,v\le1$, it is possible to analytically continue outside of this domain. To do so, we must take into account any new non-perturbative effects associated with a branch cut crossing through the integration contour as we move $u,v$ in their corresponding complex planes. More details on the analytic continuation outside of the physical domain will be given in the longer companion paper \cite{ToAppear}.
\begin{figure}[t]
\centering
\includegraphics[width=0.8\textwidth]{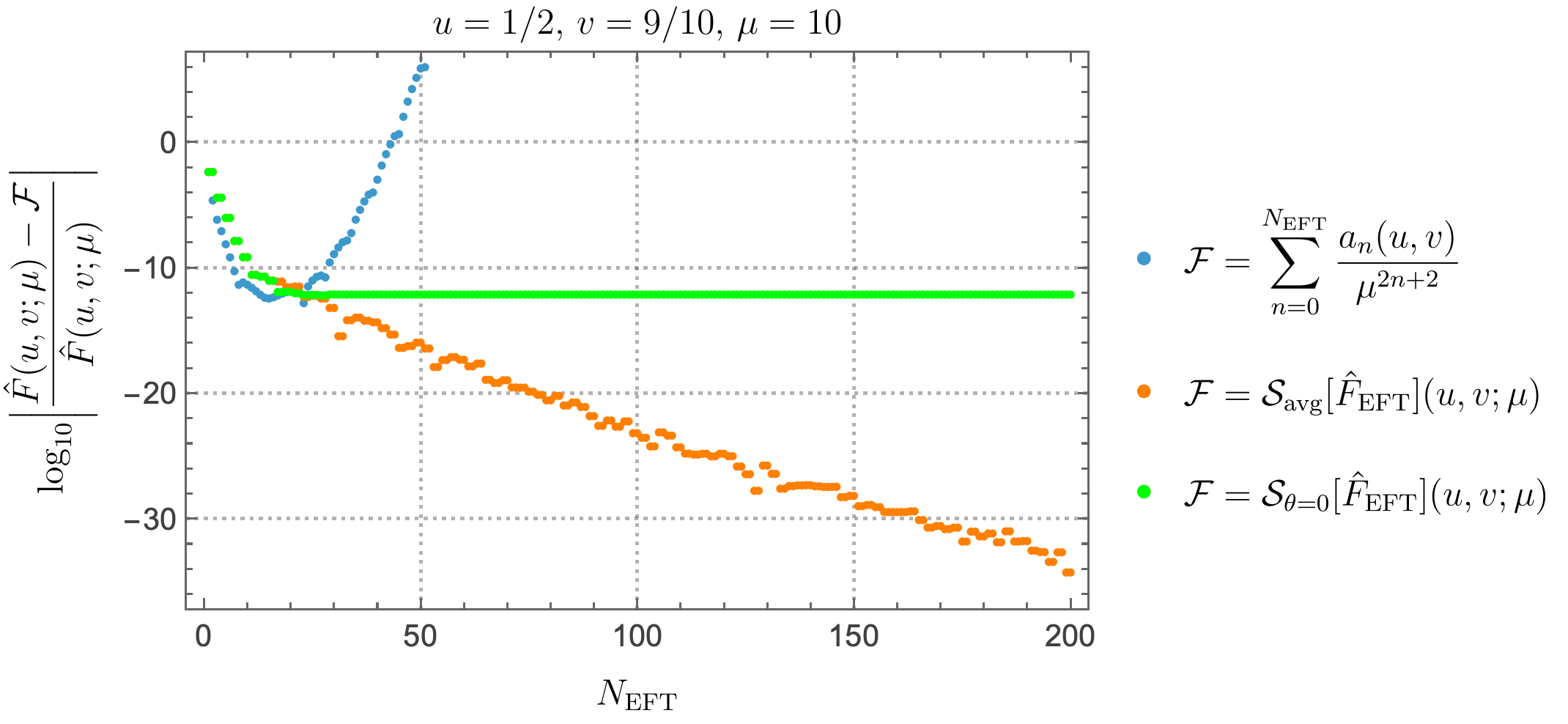}
\caption{Relative error between the dS $S$-matrix and the low-energy EFT expansion (blue), the EFT diagonal Borel-Pad\'e resummation along the imaginary axis (orange) as well as the real axis (green) as a function of the number of perturbative terms $N_{\rm EFT}$ included, at fixed $u=1/2$, $v=9/10$, and $\mu=10$. 
}
\label{fig:relerrorcomparison200}
\end{figure}

The numerical performance of our prescription is illustrated in Figure \ref{fig:relerrorcomparison200}, which compares the relative errors obtained using different contour choices with respect to the 2D de Sitter $S$-matrix, alongside the EFT approximation, as the number of perturbative terms is increased. We observe that, with the imaginary-axis prescription, the error continues to decrease as more terms are included. 

By contrast, the incorrect choice of the real axis yields no improvement over optimal truncation. This is a consequence of the fact that in rotating the contour of integration from the real axis to the correct imaginary axis contour we must include Hankel contours surrounding the branch cuts thus producing non-perturbative, exponentially suppressed contributions. The resummation along the real axis then converges to a function which differs from the 2D de Sitter $S$-matrix by exponentially suppressed terms.

We also stress that a different choice of $\beta_0$ would have produced different exponentially suppressed, non-perturbative terms. The numerical precision used in Figure \ref{fig:relerrorcomparison200} is, however, high enough that any such difference would appear as a visible discrepancy between the resummation and the conjectured physical quantity.

Our results motivate our central conjecture: the same contour prescription should reproduce the $S$-matrix data for any EFT with analogous physical constraints whose low-energy expansion is asymptotic. In other words, we conjecture that the contour dictated by the analytic structure of the Borel singularities, together with its starting point in the physical region, provides a universal prescription for extracting $S$-matrix data from asymptotic EFT expansions.

\section{Discussion}\label{sec:discussion}
In this work, we have constructed a resummation prescription for the asymptotic EFT expansion arising from the large-$\mu$ expansion of cosmological quantities. By imposing the relevant physical requirements, we select a prescription that captures non-perturbative information and yields the de Sitter $S$-matrix in two dimensions.

We expect this prescription to be robust under changes of kinematic variables and mass parametrization. Although such changes may alter the type and location of the Borel-plane singularities, we expect their corresponding non-perturbative effects, and hence the essential analytic structure underlying the prescription, to remain invariant. This motivates the conjecture that our approach extends to other processes in cosmology and beyond with analogous physical boundary conditions.

However, the prescription does depend on the boundary conditions used to select the physical resummation. In particular, the Bunch--Davies vacuum plays an essential role in our construction. If one instead considers alternative initial states, such as $\alpha$-vacua~\cite{Mottola:1985,Allen:1985,Jain:2023}, the relevant physical constraints would change and a different resummation prescription would be required.

An important feature of our result is that the resummation prescription selects the de Sitter $S$-matrix, rather than the wave function coefficient or the cosmological correlator. In fact, the latter two objects \textit{cannot} be reconstructed from the perturbative EFT expansion by any choice of integration contour in the Borel plane. 

To see this, note that reconstructing the correlator requires homogeneous contributions that are exponentially suppressed as $\sim e^{-2n\pi\mu}$, with $n\in\mathbb{N}$. However, in the physical kinematic regime, the Borel singularities lie at locations satisfying $\operatorname{Re}(t_\star)=(2n+1)\pi$, so the associated non-perturbative contributions have exponential scales of the form $\sim e^{-(2n+1)\pi\mu}$, as expected from the general relation between Borel singularities and non-perturbative sectors, see equation \eqref{eq:expterm}. These cannot reproduce the required homogeneous terms. 

Similarly, reconstructing the wave function coefficient requires additional functions of $\mu$ that cannot be generated by the resurgent structure of the perturbative EFT series alone. A conceptual understanding of why this procedure recovers the $S$-matrix but not the wave function coefficient or the correlator remains an open question.

We attribute this discrepancy to the structure of the underlying equation. Although the EFT expansions of all three physical observables are derived from the same differential equation \eqref{diffeq}, this equation does not pose the appropriate resurgence problem. The large parameter $\mu$, with respect to which we wish to apply resurgence, enters \eqref{diffeq} only as a multiplicative factor. The resurgent structure that fixes the correct homogeneous terms must therefore be found in a different class of differential or functional equations with respect to $\mu$.

There are many interesting directions to pursue as follow-up. It would be interesting to see how the large-order growth of the EFT coefficients and the associated Borel singularity structure depend on the spin of the exchanged particle, the spins and masses of the external particles, and the number of external particles~\cite{Baumann:2020dch}. More broadly, our formalism can be applied to cosmological processes whose solutions are unknown, provided the coefficients of the EFT expansion can be obtained by inverting the governing differential equations in the kinematic variables or by other means. The techniques can also be applied to cases where Lorentz invariance is broken~\cite{Cheung:2007st,Bartolo:2015qvr,Grall:2020tqc,Cheung:2026wrz}. We also expect analogous methods to apply beyond cosmology,  to theories in time-dependent backgrounds where heavy degrees of freedom are integrated out, and to thermal field theories, where thermal fluctuations can give rise to non-perturbative effects. 

Our results suggest a new way of thinking about EFTs in cosmology and other time-dependent settings: as a window into possible non-perturbative completions.
\vskip 10pt
\noindent {\bf Acknowledgments} We thank Cliff Burgess, Enrico Pajer, Carlos Duaso Pueyo, Vasudev Shyam, Matthew Walters and especially Yuanzhao Li and Zhong-Zhi Xianyu for discussions and a heads-up on their recent work \cite{Li:2026xhw}.
 DD is supported by
the Royal Society grants ICA$\backslash$R2$\backslash$242058 and IEC$\backslash$R3$\backslash$243103.
LE and GLP are supported by the ERC (NOTIMEFORCOSMO, 101126304), by Scuola Normale, and by INFN (IS GSS-Pi). GLP is also supported by the Italian Ministry of Universities and Research
(MUR) under contract 20223ANFHR (PRIN2022). 
LE presented preliminary results at the workshops: Cosmological Correlators in Taiwan 2024 and UK-Japan Workshop on Quantum Gravity 2026, as well as a seminar at the University of Edinburgh and Heriot-Watt University in 2026, and thanks the participants for questions and interesting discussion.

\appendix
\section{$S$-matrix elements, wave function coefficients, and in-in correlators}
\label{app:ds-observables}

Various cosmological objects describing the exchange process of a heavy scalar satisfy the same differential equation \eqref{diffeq}, and differ only by terms that are exponentially suppressed for momenta in the physical regime. However, the physical interpretation of the solution depends on the choice of boundary conditions. We distinguish three cases: (i) the in-in expectation value of fields at late times,  (ii) the wave function coefficient and (iii) the de Sitter $S$-matrix. The purpose of this appendix is to present the general solution to \eqref{diffeq}, explain the physical meaning of the three quantities, and show how they are related to each other.

\paragraph{The general solution and the parameter $\beta_0$.}

The particular solution to \eqref{diffeq} that satisfies the boundary conditions presented in Section \ref{sec:div} was derived in \cite{Arkani-Hamed:2018kmz}. The general solution can be expressed as a power series particular solution plus a homogeneous piece containing terms that are exponentially suppressed when $u$ and $v$ are in the physical regime,
\begin{equation}\label{fullsol}
    \hat{F}(u,v;\mu)=\begin{cases}
        \overset{\infty}{\underset{m,n=0}{\sum}} c_{m,n}(\mu)u^{2m+1}(u/v)^n+\frac{\pi}{2 \cosh(\pi \mu)}\hat{g}(u,v;\mu) & u\leq v\,, \\
        \overset{\infty}{\underset{m,n=0}{\sum}}c_{m,n}(\mu)v^{2m+1}(v/u)^n+\frac{\pi}{2 \cosh(\pi \mu)}\hat{g}(v,u;\mu) & u\geq v\,.
    \end{cases}
\end{equation} 
The series coefficients $c_{m,n}$ are given by 
\begin{equation}
     c_{m,n}(\mu)
 \coloneqq\frac{(-1)^n(n+1)_{2m}}
 {\displaystyle\prod_{j=0}^{m}
       \left[(n+\tfrac12+2j)^2+\mu^2\right]}\,.
\end{equation}
The second term in each line of \eqref{fullsol} corresponds to a solution to the homogeneous problem set by the boundary conditions and given by
\begin{align}
    \hat{g}(u,v;\mu)\coloneqq &\hat{F}_+(u;\mu)\hat{F}_-(v;\mu)-\hat{F}_-(u;\mu)\hat{F}_+(v;\mu)\nn\\&-\frac{\alpha_-(\mu)}{\alpha_+(\mu)}(\beta_0(\mu)+1)\hat{F}_+(u;\mu)\hat{F}_+(v;\mu)-\frac{\alpha_+(\mu)}{\alpha_-(\mu)}(\beta_0(\mu)-1)\hat{F}_-(u;\mu)\hat{F}_-(v;\mu)\nn\\
    &+\beta_0(\mu)[\hat{F}_-(u;\mu)\hat{F}_+(v;\mu)+\hat{F}_-(v;\mu)\hat{F}_+(u;\mu)], \label{eq:ghat}
\end{align}
with
\begin{equation}\label{eq:alphapm}
    \alpha_\pm(\mu)\coloneqq-\left(\frac{i}{2\mu}\right)^{\frac{1}{2}\pm i\mu}\frac{\Gamma(1\pm i\mu)}{\Gamma(\frac{1}{4}\pm\frac{i\mu}{2})\Gamma(\frac{3}{4}\pm\frac{i\mu}{2})},
\end{equation}
and $\hat{F}_\pm$ are the homogeneous solutions to $\eqref{diffeq}$,
\begin{align}\label{homogsol}
    \hat{F}_{\pm}(u;\mu)\coloneqq\left(\frac{iu}{2\mu}\right)^{\frac{1}{2}\pm i\mu}{}_2F_1\left(\frac{1}{4}\pm \frac{i\mu}{2},\frac{3}{4}\pm\frac{i\mu}{2},1\pm i\mu,u^2\right).
\end{align}

\paragraph{The three observables.}

The freedom in the choice of the parameter $\beta_0$ in \eqref{eq:ghat} reflects the fact that the three different physical observables discussed above all solve the same differential equation \eqref{diffeq}, hence they can only differ by a homogeneous piece. 
The corresponding choices for $\beta_0$ are summarized as follows.

In a general four-point process, the kinematics can be parametrized by six variables:  $k_i\coloneqq|\mathbf{k}_i|$, together with $s\coloneqq|\mathbf k_1+\mathbf k_2|$ and $t\coloneqq|\mathbf k_1+\mathbf k_3|$. Scale invariance reduces the number of independent variables to five. We further restrict to the $s$-channel part of the process, reducing the kinematic dependence to two dimensionless variables, $u$ and $v$.  
We introduce the following variables and parameter
\begin{equation}
u\coloneqq\frac{s}{k_1+k_2},\qquad v\coloneqq\frac{s}{k_3+k_4},\qquad
 \mu^2\coloneqq\frac{m_{\chi}^2}{H^2}-\frac{d^2}{4}>0,
 \label{eq:app-exchange-variables}
\end{equation}
where $d$ is the number of spatial dimensions and $H$ is the Hubble constant.

The \emph{in-in correlators} are expectation values in the evolved
initial state~\cite{Adshead:2009}. The final fields are integrated over, rather than fixed, with the probability weight given by the Born rule, $ |\Psi[\Phi;\tau_0]|^2 $:
\begin{equation}
 \langle O(\tau_0)\rangle
 =\frac{\int\mathcal D\Phi\,|\Psi[\Phi;\tau_0]|^2 O[\Phi]}
        {\int\mathcal D\Phi\,|\Psi[\Phi;\tau_0]|^2}.
 \label{eq:app-born-rule}
\end{equation}
For our exchange diagram, the expectation value can be denoted by $\langle\phi^4\rangle_s$, where the subscript denotes the $s$-channel part, and is related to the general solution \eqref{fullsol} by 
\begin{equation}
\langle\phi^4\rangle_s=\frac{(2\pi)^3}{16\,s\, k_1k_2k_3k_4}\delta\left(\sum_{i=1}^4\mathbf{k}_i\right) 
\times \hat{F}(u,v;\mu)\Big\vert_{ \beta_0(\mu)=\frac{1}{i\sinh(\pi\mu)}}\,.
\end{equation}

The \emph{wave function coefficients}~\cite{Benincasa:2022gtd} instead specify the amplitude for
a field configuration $\Phi$ at a time slice $\tau_0$:
\begin{equation}
\Psi[\Phi;\tau_0]=\langle\Phi,\tau_0|0\rangle_{\rm BD}\,.
\end{equation}
We use the convention
$\log\Psi=\sum_{n\geq2}\int\psi_n\Phi^n/n!$, with field indices and
momentum integrals implicit. The wave function coefficients are the $\psi_n$. The wave function retains a final-time dependence~\cite{Melville:2024ove},
\begin{equation}
 R_\mu=\frac{f^-_\mu(s\tau_0)}{f^+_\mu(s\tau_0)}
 =-e^{-\pi\mu}
   \frac{H^{(1)}_{i\mu}(-s\tau_0)}{H^{(2)}_{i\mu}(-s\tau_0)} .
 \label{eq:app-wave}
\end{equation}
For real $\mu$, this ratio oscillates as $\tau_0\to0$; the $\beta_0$ coefficient, therefore, contains a dependence on the late-time
regulator which drops out of any observable. The relation between the wave function coefficient and the general solution \eqref{fullsol} is then
\begin{equation}   \psi_{4,s}=\frac{(2\pi)^3}{16\,s\, k_1k_2k_3k_4}\delta\left(\sum_{i=1}^4\mathbf{k}_i\right)\times \hat{F}(u,v;\mu)\Big\vert_{ \beta_0(\mu)=\frac{R_\mu}{i\sinh(\pi\mu)}}\,.
\end{equation}

Lastly, the \emph{de Sitter $S$-matrix} encodes transition amplitudes between
asymptotic free-particle states \cite{Melville:2023kgd,Melville:2024ove}. In the Bunch--Davies particle basis and for two spacetime dimensions, the de Sitter $S$-matrix is related to the general solution \eqref{fullsol} via
\begin{equation}    \mathcal{S}_{0\rightarrow4}^{(D=2)}=s^{-1}\times\hat{F}(u,v;\mu)\Big\vert_{ \beta_0(\mu)=0},
\end{equation} 
where the on-shell condition in $D=2$ for asymptotic states corresponding to conformally coupled scalars with momentum of the form $k_i^\mu=(\omega_i,\mathbf{k}_i)$ is given by $\omega^2=\mathbf{k}^2$.

\paragraph{From two to four dimensions.}
There are two useful relations to uplift the $S$-matrix from two to four dimensions \cite{Melville:2024ove}.
First, for conformally coupled external particles, the physical
four-dimensional $S$-matrix can be obtained directly from the two-dimensional
$S$-matrix by replacing 
\begin{equation}
    k_j\rightarrow\omega_j=\begin{cases}
        +|\mathbf{k}_j|-i\,0, & \textrm{outgoing} \\
        -|\mathbf{k}_j|+i\,0, & \textrm{incoming}
    \end{cases}
\end{equation} and restoring external
normalizations. For four outgoing legs, this is given by
\begin{equation}
 \mathcal S_{0\to4}^{(D=4)}
 =\frac{1}{\left(\prod_{i=1}^4\sqrt{2ik_i}\right)}
   \mathcal S_{0\to4}^{(D=2)}\bigg|_{k_j\rightarrow\omega_j=|\mathbf{k}_j|} \,.
 \label{eq:app-conformal-projection}
\end{equation}
This relation applies to the tree-level contact and exchange diagrams.

Second, the four-dimensional $S$-matrix
can be generated by differentiation:
\begin{equation}
 \mathcal S_{0\rightarrow4}^{(D=4)}(u,v;\mu)
 =s(uv)^2\partial_u\partial_v\left( s^{-1}\mathcal S_{0\rightarrow4}^{(D=2)}(u,v;\mu)\right) .
 \label{eq:app-dimension-raising}
\end{equation}
This identity holds at fixed $\mu$, so
$m_{\chi,4}^2=m_{\chi,2}^2+2H^2$.
Indeed, with $\Delta_u^{(D=4)}=\Delta_u-2u\partial_u$,
\begin{equation}
 \left(\Delta_u^{(D=4)}+\mu^2+2\right)
 s^{-1}\mathcal S_{0\rightarrow4}^{(D=4)}(u,v;\mu)
 =2\left(\frac{uv}{u+v}\right)^3 ,
 \label{eq:app-raised-equation}
\end{equation}
and similarly in $v$. For general external masses, the resulting
$S$-matrix must still be evaluated with the appropriate dispersion relations.

\section{Analysis of Borel Singularity Trajectories}
\label{app:trajectories}

In this Appendix, we analyze the motion of the Borel singularities as the kinematic variable $u$ is analytically continued, while keeping $v$ fixed in the physical regime, $0<v<1$. Starting from the singularity locations given in \eqref{eq:Branches}, we determine for which values of $u\in \mathbb{C}$ a branch cut passes through the imaginary Borel axis ${\rm Re}\,t=0$.

It is useful to separate the analysis according to the magnitude of the complex variable,
\begin{equation}
u=|u|e^{i\theta},\quad -\pi\le\theta\le\pi
\end{equation}
into the two regimes $0< |u|\le1$ and $|u|>1$. The dependence of the Borel singularities on $u$ enters through the quantity $\log s_u$. More explicitly, the real part of their locations takes the form
\begin{equation}\label{eq:Rets}
\operatorname{Re}t_\star=(2n+1)\pi\pm\arg s_u.
\end{equation}
We shall be using the convention $-\pi\leq\arg \,z\leq \pi$. Hence, we deduce from \eqref{eq:Rets} that ${\rm Re}\, t_\star$ can vanish only when $\arg s_u=\pm\pi$ and for $n=-1,0$.

\subsubsection*{Case I: $0< |u|\leq 1$}
It is useful to express the trajectory analysis in terms of $s_u$, defined in \eqref{eq:quant}. The relation between $s_u$ and $u$ can be inverted to give
\begin{equation}\label{eq:inversion}
     u=\frac{2s_u}{1+s_u^2}.
\end{equation}

If $s_u$ lies on the negative real axis, this relation implies that $u$ is also real and negative.
 Equivalently, if $u$ does not lie on the negative real axis, then $s_u$ does not lie on the negative real axis. Thus, for $-\pi<\theta<\pi$, $s_u$ does not cross the branch cut of the logarithm, and the phase of $s_u$ remains on the principal branch,
\begin{equation}
-\pi<\arg s_u<\pi.
\end{equation}
Since the real part of the Borel singularities \eqref{eq:Rets} can vanish only when $\arg s_u=\pm\pi$, no singularity can reach $\operatorname{Re}t_\star=0$ away from the negative real axis. Therefore, the only possible crossing occurs on the boundary $\theta=\pm\pi$.

Now approaching the negative real axis with $u$ from above with
\begin{equation}\label{eq:above}
\theta=\pi-\epsilon,\qquad
\epsilon\rightarrow0^+,
\end{equation}
we find
\begin{equation}\label{eq:logsu}
\log s_u
=
\log s_{|u|}
-i\pi+i\epsilon f_u,
\end{equation}
where
\begin{equation}
f_u=\frac{1}{\sqrt{1-|u|^2}}.
\end{equation}
The locations of the branch points from equation \eqref{eq:Branches} then become
\begin{equation}\label{eq:Branchesabove}
t_\star = (2n+1)\pi+ \begin{cases}-\pi
+\epsilon f_u-i\log(s_{|u|} s_v)\,, \\
\phantom{-}\pi
-\epsilon f_u-i\log(\frac{1}{s_{|u|} s_v})\,,\\
-\pi
+\epsilon f_u-i\log(\frac{s_{|u|}}{s_v})\,,\\
\phantom{-}\pi
-\epsilon f_u-i\log(\frac{s_v}{s_{|u|}})\,.
\end{cases}
\end{equation}
From this, we can identify the branch points whose real parts vanish in the limit $\epsilon\to0^+$, reproducing~\eqref{eq:revanishabove}.

Next, approaching the negative real axis from below with 
\begin{equation}\label{eq:below}
\theta=-\pi+\epsilon,\qquad
\epsilon\rightarrow0^+,
\end{equation}
we find
\begin{equation}\label{eq:logsubelow}
\log s_u
=
\log s_{|u|}
+i\pi-i\epsilon f_u,
\end{equation}
The locations of the branch points from equation \eqref{eq:Branches} then become
\begin{equation}\label{eq:Branchesbelow}
t_\star = (2n+1)\pi+ \begin{cases}\phantom{-}\pi
-\epsilon f_u-i\log(s_{|u|} s_v)\,, \\
-\pi
+\epsilon f_u-i\log(\frac{1}{s_{|u|} s_v})\,,\\
\phantom{-}\pi
-\epsilon f_u-i\log(\frac{s_{|u|}}{s_v})\,,\\
-\pi
+\epsilon f_u-i\log(\frac{s_v}{s_{|u|}})\,,
\end{cases}
\end{equation}
from which we get \eqref{eq:revanishbelow}. Comparing \eqref{eq:Branchesabove} and \eqref{eq:Branchesbelow}, we see explicitly that analytic continuation around the negative real axis exchanges which Borel singularities approach the imaginary axis.

As noted in the main text, the case $|u|=1$ has to be treated slightly differently since $f_u$ diverges there. However, from the locations of the branch points given in  \eqref{eq:revanishabove1}-\eqref{eq:revanishbelow1} we deduce that even for $|u|=1$ Borel singularities approach the imaginary axis only for $u = e^{i(\pi -\epsilon)}$ and  $u = e^{i(-\pi +\epsilon)}$ with $\epsilon \to 0^+$.

Thus, within the unit disk, the only values of $u$ for which Borel singularities can reach the imaginary axis $\operatorname{Re}t_\star=0$ lie on the interval $-1\leq u<0$.

\subsubsection*{Case II: $|u|>1$}
For $|u|>1$, the same inversion relation \eqref{eq:inversion}
shows that for the same reason as with $0<|u|<1$, no additional branch crossing can occur away from the negative real $u$-axis. It therefore remains only to examine the boundary $\theta=\pm\pi$.

Once again approaching the negative real axis with $u$ from above as in \eqref{eq:above} we find
\begin{equation}
    \log s_u=i\left(-\pi+\arctan\sqrt{|u|^2-1}\right)+\frac{\epsilon}{\sqrt{|u|^2-1}}.
\end{equation}
The locations of the branch points from  \eqref{eq:Branches} then become
\begin{equation}\label{eq:BranchesaboveII}
t_\star = (2n+1)\pi+ \begin{cases}-\pi+\arctan\sqrt{|u|^2-1}
-\frac{i\epsilon}{\sqrt{|u|^2-1}}-i\log s_v\,, \\
\phantom{-}\pi-\arctan\sqrt{|u|^2-1}
+\frac{i\epsilon}{\sqrt{|u|^2-1}}-i\log(\frac{1}{ s_v})\,,\\
-\pi+\arctan\sqrt{|u|^2-1}
-\frac{i\epsilon}{\sqrt{|u|^2-1}}-i\log(\frac{1}{s_v})\,,\\
\phantom{-}\pi-\arctan\sqrt{|u|^2-1}
+\frac{i\epsilon}{\sqrt{|u|^2-1}}-i\log s_v\,,
\end{cases}
\end{equation}
From this, we see that since for $|u|>1$

\begin{equation}
0<\arctan\sqrt{|u|^2-1}<\frac{\pi}{2},
\end{equation}
none of the above expressions can have a vanishing real part for any integer $n$. 

Similarly approaching from below with \eqref{eq:below}, we find
\begin{equation}
    \log s_u=i\left(\pi-\arctan\sqrt{|u|^2-1}\right)+\frac{\epsilon}{\sqrt{|u|^2-1}},
\end{equation}
and the locations of the Borel singularities become
\begin{equation}\label{eq:BranchesbelowII}
t_\star = (2n+1)\pi+ \begin{cases}\phantom{-}\pi-\arctan\sqrt{|u|^2-1}
-\frac{i\epsilon}{\sqrt{|u|^2-1}}-i\log s_v\,, \\
-\pi+\arctan\sqrt{|u|^2-1}
+\frac{i\epsilon}{\sqrt{|u|^2-1}}-i\log(\frac{1}{ s_v})\,,\\
\phantom{-}\pi-\arctan\sqrt{|u|^2-1}
-\frac{i\epsilon}{\sqrt{|u|^2-1}}-i\log(\frac{1}{s_v})\,,\\
-\pi+\arctan\sqrt{|u|^2-1}
+\frac{i\epsilon}{\sqrt{|u|^2-1}}-i\log s_v\,.
\end{cases}
\end{equation}
Again, none of the real parts can vanish for any integer $n$. Hence, no Borel singularity reaches the imaginary axis for $|u|>1$.

Combining the two regions, we conclude that the real parts of the Borel singularities can vanish only for
\begin{equation}
\boxed{-1\leq u\leq0}\,.
\end{equation}

\section{Harmonic Oscillator Resummation}
\label{app:HOresummation}

In this appendix, we want to analyze the resummation prescription  \eqref{eq:contour} specialized to the harmonic oscillator limit,~i.e. we wish to study the analytic properties in the $\xi_0$ variable for the averaged resummation
\begin{equation}
\mathcal{S}_{\rm avg}[F_{\textrm{EFT}}](\xi_0;\mu) = \frac{1}{2\mu}\left(\int_0^{i \infty} e^{-t\mu} B(\xi_0;t) \,{\rm d}t+\int_0^{-i \infty} e^{-t\mu} B(\xi_0;t) \,{\rm d}t\right)\,,\label{eq:SHOavg}
\end{equation}
where we restrict the domain for $\xi_0$ to the physical regime $0<\xi_0<\infty$.
In the companion paper \cite{ToAppear} we shall show that when we analytically continue \eqref{eq:SHOavg} outside of the physical domain some singularities of the integrand will hit the contour of integration.
As a consequence, \eqref{eq:SHOavg} will have to be modified by adding non-perturbative terms which will capture precisely the discontinuity properties of the desired, exact physical solution.

For convenience, we rewrite here the Borel transform \eqref{eq:Btwoterm} for the harmonic oscillator limit,
\begin{align}
B(\xi;t)
=\frac12\left[\frac{\xi e^{\frac{it}{2}}}{1+\xi e^{it}}+\frac{\xi e^{-\frac{it}{2}}}{1+\xi e^{-it}}\right]\,,
\label{eq:HOBorel}
\end{align}
whose poles are located at 
\begin{equation}
t_{\star,n}^\pm(\xi) \coloneqq  (2n+1)\pi \pm i \log(\xi)\,,\qquad n\in\mathbb Z\,.\label{eq:SHOpoles}
\end{equation}
We shall here restrict our analysis to the case $0<\xi_0<1$, where the $t_{\star,n}^+$ poles lie below the real axis, while the $t_{\star,n}^-$ poles lie above it. However, as will be shown in the companion paper \cite{ToAppear}, a similar analysis can be carried out for $\xi_0>1$.

\begin{figure}[t]
  \centering
  \begin{subfigure}[t]{0.49\textwidth}
    \centering
    \includegraphics[width=\textwidth]{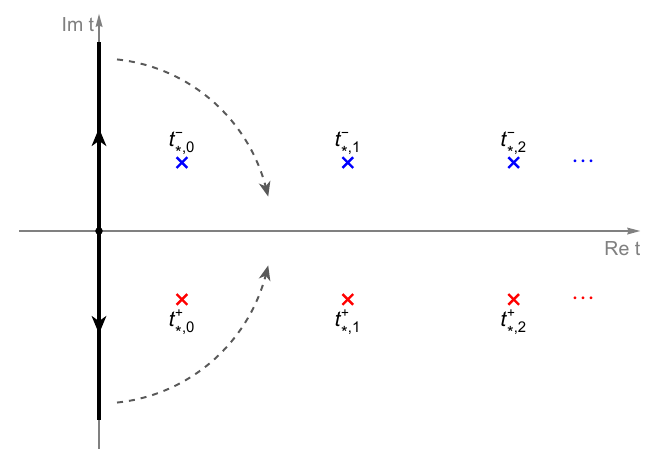}
    \caption{Integration contours for the averaged resummation along $\arg(t)=\pm\frac{\pi}{2}$.}
    \label{fig:SHOavg}
  \end{subfigure}\hfill
  \begin{subfigure}[t]{0.49\textwidth}
    \centering
    \includegraphics[width=\textwidth]{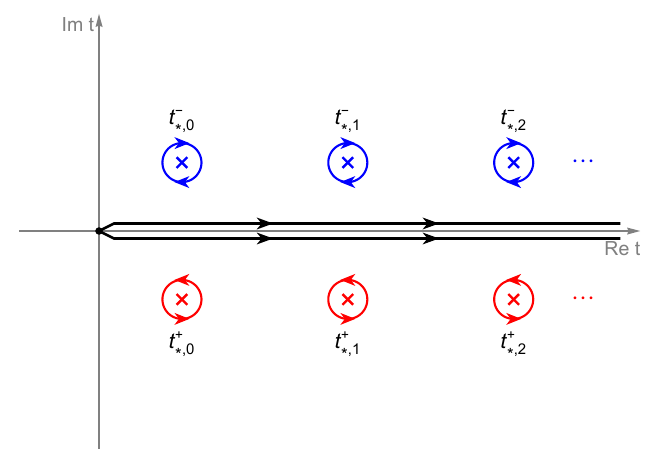}
    \caption{Integration contour along $\arg(t){=}0$, plus clockwise/counter-clockwise circles around
      $t^-_{\star,n}$ and $t^+_{\star,n}$ respectively.}
    \label{fig:SHOreal}
  \end{subfigure}
  \caption{Change in integration contour from the imaginary axis to the real line.}
  \label{fig:SHOcontours}
\end{figure}

As illustrated in Figure \ref{fig:SHOavg}, 
the original contour of integration in \eqref{eq:SHOavg} is the average of the two rays along the positive and
negative imaginary axes, each entering with weight $1/2$. Rotating the upper
ray clockwise onto the positive real axis sweeps across the poles
$t^-_{\star,n}$, which are therefore encircled clockwise. Similarly, the lower ray sweeps across the poles $t^+_{\star,n}$, encircled
counter-clockwise. Each pole contribution therefore keeps the weight $1/2$ of
the ray it came from. The two rotated rays both end on the
positive real axis, drawn as two separate lines in
Figure~\ref{fig:SHOreal} for clarity, so that we are left with an integral along the positive real axis with weight $1/2+1/2=1$.

We wish to express the harmonic oscillator resummation in terms of the particular solution $F_P$ in \eqref{eq:FP}. Starting with the integral representation of the hypergeometric functions \cite[eq.~15.6.1]{NIST:DLMF}, valid for $\xi\notin(-\infty,-1]$, we have
\begin{equation}\label{eq:2F1intrep}
{}_2F_1\!\Big(\tfrac12\pm i\mu,1;\tfrac32\pm i\mu\,\Big|-\xi\Big)
=\frac{1\pm 2i\mu}{2}\int_0^1\big[(1+x\xi)(1-x)\big]^{-\frac{1}{2}}\left[\frac{1-x}{1+x\xi}\right]^{\pm i\mu}dx\,.
\end{equation} 
Using this representation, we find that the particular solution $F_P$ defined in \eqref{eq:FP} can be written as the directional Borel resummation for the  perturbative EFT expansion \eqref{eq:FPert} along the positive real axis when $0<\xi_0<1$, that is
\begin{equation}
    F_P(\xi_0;\mu)= \mathcal{S}_0[F_{\rm EFT}](\xi_0;\mu) = \frac{1}{\mu}\int_0^{ \infty} e^{-t\mu} B(\xi_0;t) \,{\rm d}t.\label{eq:S0Fp}
\end{equation}
Thus, the integral along the positive real axis reproduces the particular solution $F_P(\xi_0;\mu)$. Note that this  expression for $F_P$ is only guaranteed to hold for the range $0<\xi_0<1$. For values outside this range, Borel singularities do cross the real axis and non-perturbative terms must be added to \eqref{eq:S0Fp}. 

It remains to evaluate the contributions from the residues of the poles. Summing over all residues $n\ge0$ gives,
\begin{align}
    \frac{i \pi}{\mu}
\sum_{n\ge0}\left(\operatorname{Res}_{t=t^+_{\star,n}}\big[e^{-t\mu}B(\xi_0;t)\big]{-}\operatorname{Res}_{t=t^-_{\star,n}}\big[e^{-t\mu}B(\xi_0;t)\big]
\right){=}\frac{i\pi\,\left(\xi_0^{\frac{1}{2}+i\mu}-\xi_0^{\frac{1}{2}-i\mu}\right)}{2\mu(e^{\pi\mu}+e^{-\pi\mu})}\,.
\end{align}
Combining the integral and residue contributions, we obtain
\begin{equation}
    \mathcal{S}_{\rm avg}[F_{\textrm{EFT}}](\xi_0;\mu)=F_P(\xi_0;\mu)+\frac{i\pi}{4\mu\cosh{\pi\mu}}\left(\xi_0^{\frac{1}{2}+i\mu}-\xi_0^{\frac{1}{2}-i\mu}\right),
\end{equation}
 which holds for all $\xi_0$ in the physical regime $0<\xi_0<\infty$. 

\bibliographystyle{JHEP}
\bibliography{draft.bib}
\end{document}